\documentclass[a4paper,fleqn,final]{cas-dc}

\usepackage{amsmath,amsfonts}
\usepackage{algorithmic}
\usepackage{algorithm}
\usepackage{array}
\usepackage[caption=false,font=scriptsize,labelfont=sf,textfont=sf]{subfig}
\usepackage{textcomp}
\usepackage{url}
\usepackage{verbatim}
\usepackage{graphicx}
\usepackage{cite}
\usepackage{xcolor}
\usepackage{multirow}
\usepackage{tikz}
\usepackage{booktabs}
\usepackage{kotex}
\usepackage{soul}
\usepackage{makecell}
\usepackage{wasysym}
\hypersetup{breaklinks=true}
\usepackage{xurl}

\newcommand{\hmseo}[1]{\textcolor{black}{#1}}

\newcommand{\ourtool}{\textsc{SecTracer}}
\newcommand*{\tpc}{\texttt{2PC}}

\newcommand*\circled[1]{\tikz[baseline=(char.base)]{
            \node[shape=circle,draw,inner sep=0.5pt] (char) {#1};}}

\usepackage[authoryear,longnamesfirst]{natbib}

\def\tsc#1{\csdef{#1}{\textsc{\lowercase{#1}}\xspace}}
\tsc{WGM}
\tsc{QE}
\tsc{EP}
\tsc{PMS}
\tsc{BEC}
\tsc{DE}

\newcommand{\fullcircle}{$\CIRCLE$}
\newcommand{\halfcircle}{$\LEFTcircle$}
\newcommand{\emptycircle}{$\Circle$}



\begin{document}
\let\WriteBookmarks\relax
\def\floatpagepagefraction{1}
\def\textpagefraction{.001}

\shorttitle{\ourtool: A Framework for Uncovering the Root Causes of Network Intrusions via Security Provenance}

\shortauthors{Lee et~al.}

\title [mode = title]{\ourtool: A Framework for Uncovering the Root Causes of Network Intrusions via Security Provenance}                     
\tnotemark[1]
\tnotetext[1]{This is the accepted manuscript of an article accepted for publication in \textit{Computers \& Security}.}

\author[1]{Seunghyeon Lee}
\ead{sl@s2w.inc}
\fnmark[1]
\author[2]{Hyunmin Seo}
\ead{shm9574@kaist.ac.kr}
\fnmark[1]

\author[3]{Hwanjo Heo}
\ead{hwanjo@etri.re.kr}
\author[4]{Anduo Wang}
\ead{adw@temple.edu}
\author[2]{Seungwon Shin}[]
\ead{claude@kaist.ac.kr}
\author[5]{Jinwoo Kim}[orcid=0000-0003-1303-8668]
\ead{jinwookim@kw.ac.kr}
\cormark[1]

\fntext[1]{Co-first authors}

\affiliation[1]{organization={S2W},
addressline={Pangyoyeok-ro 192beon-gil}, 
city={Seongnam-si},
postcode={13524}, 
country={Republic of Korea}}

\affiliation[2]{organization={KAIST},
addressline={291 Daehak-ro, Yuseong-gu}, 
city={Daejeon},
postcode={34141}, 
country={Republic of Korea}}

\affiliation[3]{organization={Electronics and Telecommunications Research Institute (ETRI)},
    city={218 Gajeong-ro, Yuseong-gu},
    city={Daejeon},
    postcode={34129}, 
    country={Republic of Korea}
}

\affiliation[4]{organization={Computer and Information Science, Temple University},
            address line ={SERC 342, N 12th St}, 
            city={Philadelphia},
            state=PA,
            postcode={19122}, 
            country={USA}
}

\affiliation[5]{organization={School of Software, Kwangwoon University},
            addressline={20 Kwangwoon-ro, Nowon-gu}, 
            city={Seoul},
            postcode={01897}, 
            country={Republic of Korea}
}

\cortext[cor1]{Corresponding author}

\begin{abstract}
Modern enterprise networks comprise diverse and heterogeneous systems that support a wide range of services, making it challenging for administrators to track and analyze sophisticated attacks such as advanced persistent threats (APTs), which often exploit multiple vectors. To address this challenge, we introduce the concept of \emph{network-level security provenance}, which enables the systematic establishment of causal relationships across hosts at the network level, facilitating the accurate identification of the root causes of security incidents. Building on this concept, we present \ourtool{} as a framework for a network-wide provenance analysis. \ourtool{} offers three main contributions: (i)~comprehensive and efficient forensic data collection in enterprise networks via software-defined networking (SDN), (ii)~reconstruction of attack histories through provenance graphs to provide a clear and interpretable view of intrusions, and (iii)~proactive attack prediction using probabilistic models. We evaluated the effectiveness and efficiency of \ourtool{} through a real-world APT simulation, demonstrating its capability to enhance threat mitigation while introducing less than 1\% network throughput overhead and negligible latency impact.
\end{abstract}


\begin{highlights}

\item We introduce network-level security provenance to analyze multihost intrusion behaviors.

\item \ourtool{}, built on SDN, uses probabilistic logic to automate root cause analysis of APTs.

\item Our approach enables accurate attack forensics with low overhead and no system modifications.

\end{highlights}


\maketitle

\section{Introduction}
With the advent of several key research innovations (e.g., network virtualization and mobility management), enterprise networks are actively evolving into the next generation, enabling them to support a wider range of diverse and complex services (e.g., e-commerce and e-voting). However, this transformation not only brings advancements but also introduces a critical challenge: network intrusions. For example, several \emph{advanced persistent threats (APTs)} targeting critical assets within enterprise networks have emerged. These threats often compromise reachable hosts, using them as entry points to infiltrate the network and steal or manipulate critical assets, a technique known as lateral movement or stepping-stone attacks~\cite{zhang2000detecting}. 

Network administrators are aware of these threats and have made considerable efforts to reduce network attack surfaces. To this end, they deployed a variety of security mechanisms (e.g., network intrusion detection systems and firewalls) to detect specific attack attempts~\cite{zuech2015intrusion, bhuyan2014network} and even outsourced security services to external providers~\cite{crowdstrike-mssp}. However, despite these efforts, sophisticated attacks persist and often bypass traditional defenses. Therefore, administrators frequently fail to trace the origins and progression of these attacks (e.g., identifying the initial entry point), leaving networks vulnerable to future unforeseen threats.

To address this challenge, many \emph{provenance-based intrusion detection systems (PIDS)} have been proposed \cite{pohly2012hi, gehani2012spade, bates2015trustworthy, pasquier2017practical, chen2021clarion, datta2022alastor, tabiban2022provtalk, irshad2021trace, dong2023we, sekar2024eaudit, hossain2017sleuth, milajerdi2019holmes, hossain2020combating, hassan2020tactical, milajerdi2019poirot, hassan2019nodoze, han2020unicorn, wang2020you, alsaheel2021atlas, wang2022threatrace, zengy2022shadewatcher, yang2023prographer, jia2024magic, goyal2024r, cheng2024kairos, rehman2024flash}. These systems collect data on system and network activities, construct provenance graphs, and analyze the root causes of security incidents. However, they have several limitations. First, most existing PIDS focus on collecting and analyzing host-level provenances, making it difficult to capture and establish causal relationships across multiple hosts. Second, their monitoring coverage is limited because achieving comprehensive visibility requires the deployment of agents on all hosts, which is both costly and operationally burdensome. Third, provenance graphs often contain a vast number of vertices and edges, making manual investigation highly time-consuming and labor-intensive. We suggest that overcoming these limitations is crucial for effective root cause analysis and forensic investigation of complex network intrusions in enterprise networks.

In this study, we introduced \ourtool{}, a novel framework for comprehensive provenance-based intrusion detection through network-wide monitoring. \ourtool{} provides administrators with clear visibility into network intrusions, including their origin and progression, by continuously monitoring and collecting relevant network and security events to enable the reconstruction, analysis, and mitigation of advanced threats. To achieve this, we introduced the concept of \textit{network-level security provenance}, an extension of traditional provenance that offers visibility into how specific network events are created, modified, and evolve, facilitating the efficient tracking of changes and precise identification of intrusion root causes. By leveraging network-level security provenance, \ourtool{} uncovers the origins of enterprise security incidents, providing deeper insights into attack paths and strengthening network defense mechanisms.

At a high level, {\ourtool} collects security and related audit data (e.g., network events) by utilizing software-defined networking (SDN). It considers packets and network snapshots---including network topology, security policies, configurations, and forwarding rules---as primitive provenance data and captures them in chronological order to provide sufficient forensic information to reconstruct the attack history. Using the collected data, {\ourtool} constructs a \emph{network-level security provenance graph} to represent the operational flow of attacks across hosts, illustrating not only the causal relationships of security events with their origins but also the forensic details that characterize attack behaviors. To generate the provenance graph, {\ourtool} begins its analysis from a specific event or host defined by the network administrator, and performs a causality analysis to uncover the relationships between discrete security diagnostic results.

To streamline root cause analysis from complex provenance graphs, \ourtool{} leverages probabilistic models~\cite{psl}. It extracts information related to network security issues from a security provenance graph and structures it into known relationships using first-order logic. By applying predefined root cause analysis (RCA) rules, \ourtool{} models the attacker behavior and prioritizes various root cause insights for each subject. Finally, it exports the constructed security provenance graph, enhancing administrators’ ability to analyze suspicious activities with maximum security visibility. This approach significantly reduces the overhead of manual security data analysis and minimizes the risk of overlooking critical information, such as the starting point of an APT attack. \\

\noindent\textbf{Contributions.} The key contributions of this study are as follows:  
\begin{itemize}  

    \item We proposed \emph{network-level security provenance} enriched with diverse features collected from an enterprise network. This is enabled by SDN, which provides centralized control and fine-grained traffic monitoring.

    \item We designed a method for performing RCA on complex provenance graphs by leveraging probabilistic soft logic (PSL), which automatically identifies intricate attack patterns based on predefined rules.
    
    \item We developed a prototype system, {\ourtool}, which implements network-level security provenance in modern enterprise networks, efficiently collecting comprehensive security and network audit data without requiring modifications to existing infrastructure or components.  

    \item We evaluated the effectiveness of {\ourtool} by demonstrating its capability to trace the root causes of intrusion scenarios (i.e., an APT attack and botnet propagation) and assess its efficiency by measuring system overhead and performance.  

\end{itemize}

The remainder of this paper is organized as follows:
Section~\ref{s:prob} details the problem statement, including a motivating example, our research goals, and the threat model.
Section~\ref{s:related} discusses the related studies on provenance-based intrusion detection.
The high-level design and architecture of {\ourtool} are presented in Section~\ref{s:design}.
Subsequently, we describe its three core components in detail: provenance collection (Section~\ref{s:design-collector}), provenance analysis (Section~\ref{s:design-analyzer}), and RCA (Section~\ref{s:design-rca}).
Section~\ref{s:eval} presents a comprehensive evaluation of the proposed framework's effectiveness and performance.
Finally, Section~\ref{s:discussion} discusses limitations and future work.


\section{Problem Statement}
\label{s:prob}

\subsection{Motivating Example}
\label{s:prob-motiv}

\begin{figure}
    \centering
    \includegraphics[width=1\linewidth]{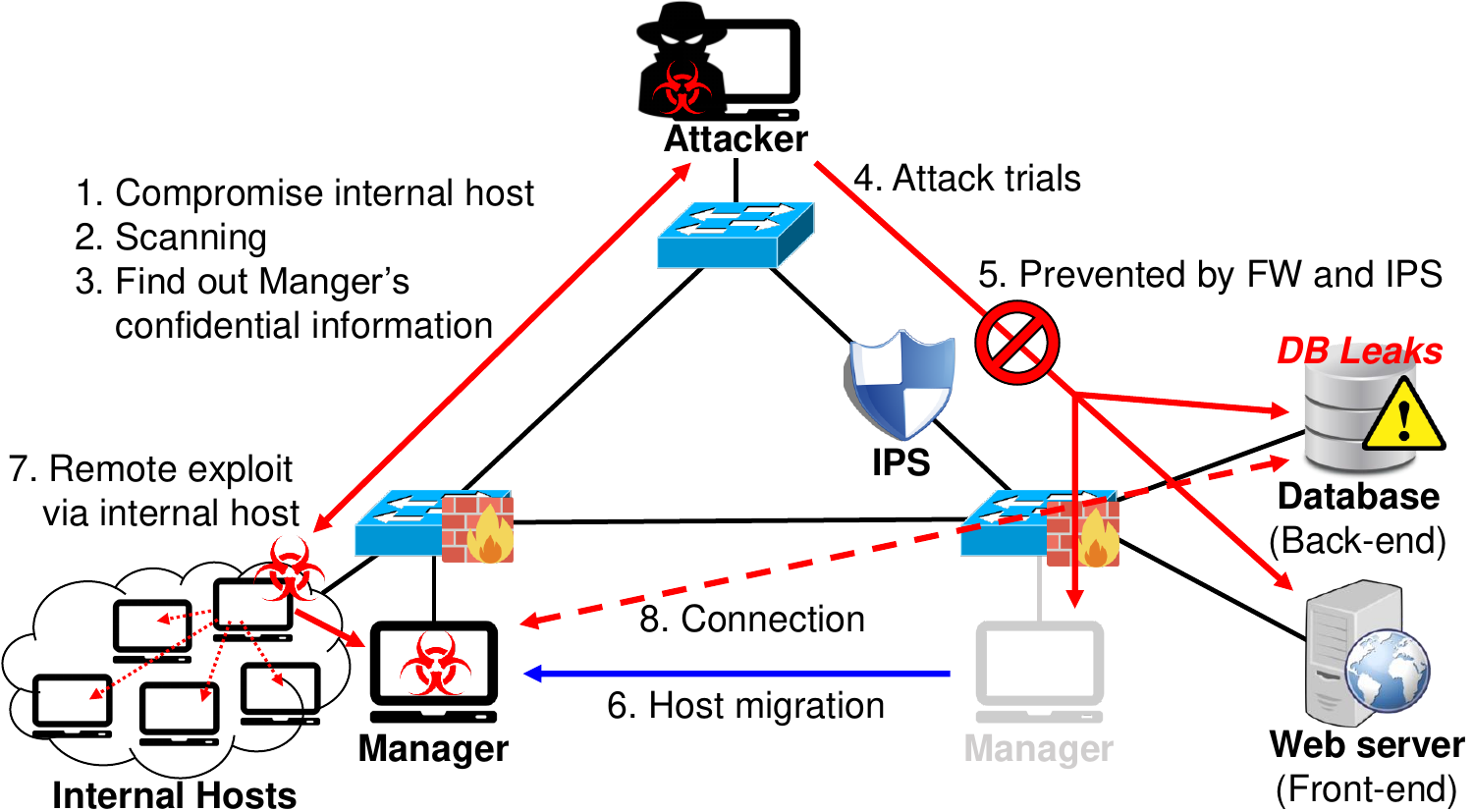}
    \caption{Example scenario where an attacker aims to steal confidential information through the manager's machine via an APT~\cite{zhang2000detecting}. The dotted line indicates that the attacker controls compromised victims.}
    \label{f:use_case_apt}
\end{figure}

\begin{table*}[!t]
\scriptsize
\caption{
Mapping of scenario steps to the MITRE ATT\&CK framework, based on an attack scenario designed to reflect realistic conditions of an APT incident.}
\begin{center}
\begin{tabular}{c|l|l}
\bf{Step} & \bf{MITRE ATT\&CK Tactic} & \bf{Technique ID \& Name} \\
\hline 
\hline 
\multirow{2}{1em}{\circled{1}} & \multirow{2}{20em}{Initial Access} & T1190 - Exploit Public-Facing Application \\ 
& & T1566 - Phishing \\ \hline 
\multirow{2}{1em}{\circled{2}} & \multirow{2}{20em}{Discovery} & T1046 – Network Service Discovery \\
& & T1018 – Remote System Discovery\\ \hline 
\multirow{2}{1em}{\circled{3}} & \multirow{2}{20em}{Credential Access / Discovery} & T1083 – File and Directory Discovery \\
& & T1555 – Credentials from Password Stores \\ \hline 
\multirow{2}{1em}{\circled{4}} & \multirow{2}{20em}{Lateral Movement / Privilege Escalation} & T1021 – Remote Services \\
& & T1078 – Valid Accounts \\ \hline 
\circled{5} & Defense Evasion (\textit{Prevented}) & T1562 – Impair Defenses (\textit{attempt failed}) \\ \hline 
\multirow{2}{1em}{\circled{6}} & \multirow{2}{20em}{Defense Evasion} & T1078 – Valid Accounts \\
& & T1562 – Impair Defenses \\ \hline 
\multirow{2}{1em}{\circled{7}} & \multirow{2}{20em}{Lateral Movement} & T1021 – Remote Services \\
& & T1550 – Use Alternate Authentication Material \\ \hline 
\multirow{3}{1em}{\circled{8}} & \multirow{3}{20em}{Privilege Escalation / Impact / Exfiltration} & T1068 – Exploitation for Privilege Escalation \\
& & T1005 – Data from Local System \\
& & T1041 – Exfiltration over C2 Channel \\
\end{tabular}
\end{center}

\label{t:scenario_mapping}
\end{table*}

Figure~\ref{f:use_case_apt} illustrates a typical APT scenario targeting an enterprise network. In this scenario, the attacker compromises the database and web servers. \circled{1}~\textbf{(Initial Access)} The attacker first attempts to directly attack an internal host and successfully gains access. \circled{2}~\textbf{(Discovery and Reconnaissance)} Next, the attacker conducts port scanning on other internal hosts to gather intelligence about the network. \circled{3}~\textbf{(Enumeration and Credential Access)} After multiple scanning attempts, the attacker uncovers details about the network infrastructure. \circled{4}~\textbf{(Lateral Movement and Privilege Escalation Attempts)} Using this information, the attacker attempts to directly compromise the database server, the web server, and the manager’s  machine. \circled{5}~\textbf{(Defense Evasion and Access Control)} This initial attempt is blocked by the intrusion prevention system (IPS) and firewall rules configured by the administrator, which restrict remote access to critical assets. At this stage, the internal systems remain well-protected. \circled{6}~\textbf{(Configuration Drift and Security Gap Exploitation)} However, after migrating the manager machine to another switch, the administrator neglects to update security configurations. \circled{7}~\textbf{(Exploitation and Lateral Movement)} Exploiting this misconfiguration, the attacker gains access to the internal host and subsequently moves laterally to the manager machine. \circled{8}~\textbf{(Privilege Escalation, Impact, and Data Exfiltration)} After attacking the manager machine,  the attacker escalates privileges, breaching the database server and gaining access to sensitive records.
To enhance the relevance and clarity of the scenario, each step of the attack was mapped onto the MITRE ATT \&CK~\cite{mitre} framework, as summarized in Table~\ref{t:scenario_mapping}.

Although all the aforementioned attack attempts could be prevented, either by maintaining the system up-to-date (e.g., patching the kernel) or by deploying advanced security measures, the primary challenge for administrators lies in the vast number of possible attack scenarios. The sheer size of an attack surface makes it extremely difficult to anticipate and mitigate potential threats. Furthermore, when an attack occurs, administrators may lack the necessary tools to detect the theft of confidential information. Security defense mechanisms may subsequently flag unusual database access patterns or document leaks; however, linking these alerts to the full sequence of attack events remains a significant challenge. In other words, although security mechanisms can capture observable signs of attack attempts, they often fail to explain their root causes, such as \emph{host migration}, or correlate historical events that combinedly constitute a successful attack. In this context, no clear or comprehensive solution exists.

\subsection{Research Goal and Limitations}
As demonstrated in the aforementioned example, understanding how an attacker exfiltrates confidential information is extremely challenging from an administrator's perspective. Therefore, our goal is to systematically analyze such complex attacks using {\it RCA}, a method for identifying the underlying origins of a problem or event. In the context of security, RCA involves a structured sequence of security diagnostics, including (i) threat recognition and impact assessment, (ii) investigation to trace the origins of security incidents, and (iii) security enhancement by addressing root causes. However, achieving these objectives in current enterprise networks is hindered by three fundamental limitations (L1--L3):\

\noindent\textbf{L1. Limited Monitoring Coverage.} To minimize the deployment overhead, many enterprise administrators position security devices at network choke points~\cite{fayaz2015bohatei,lee2017athena}, balancing concerns such as performance degradation (e.g., deep packet inspection) and network complexity (e.g., flow management). However, selective security monitoring also introduces critical vulnerabilities. As illustrated in Figure~\ref{f:use_case_apt}, key security events, such as \circled{1}, \circled{2}, and \circled{7}, are essential for detecting and preventing information leakage attacks. However, because of the attacker's ability to bypass the coverage of the IPS and firewalls, these crucial events may go unnoticed by administrators. Administrators can enhance the monitoring coverage by deploying security agents across all hosts. However, these agents continuously collect additional security-related data, which imposes resource overhead on the host and potentially degrades system performance~\cite{dong2023we, sekar2024eaudit}. Furthermore, this method is inherently limited to scenarios in which users operate devices without a preinstalled security agent because the behavior of such devices remains undetected~\cite{Zahadat2015BYOD}. Consequently, detecting security incidents at the network level is crucial because it enables the identification of threats irrespective of host-specific configurations and provides a network-wide perspective on potential security breaches. \\

\noindent\textbf{L2. Limited Network-wide Causality Analysis.}
Attackers often employ various techniques to execute multiple attack attempts to achieve their objectives. Although traditional attack detection systems, such as security information and event management (SIEM) and intrusion detection systems (IDS), can effectively identify individual attack attempts, they are limited in determining the root cause. To overcome this limitation, PIDS have been studied extensively ~\cite{pohly2012hi, gehani2012spade, bates2015trustworthy, chen2021clarion, datta2022alastor, tabiban2022provtalk}. These systems construct provenance graphs by capturing system events, such as process creation, file modifications, and network communications, offering a comprehensive view of malicious activities. The provenance graph enables a fine-grained causality analysis, allowing administrators to trace attack chains, analyze data dependencies, and pinpoint root causes. However, the existing PIDS face challenges in detecting attacks that span multiple hosts. The lack of network-wide correlation prevents them from reconstructing multihost attack paths and effectively detecting lateral movements in distributed network environments. \\

\noindent\textbf{L3. Overwhelming Security Alerts.}
Administrators must prioritize security alerts based on their severity and impact on enterprise infrastructure. However, according to Trend Micro's threat reports~\cite{trend-triage}, 70\% of security teams are emotionally overwhelmed by security alert volumes, which makes it difficult to triage all potential threats. Large and complex enterprise environments generate enormous volumes of alerts, which often lead to alert fatigue. Despite existing efforts to prioritize security events~\cite{snort-class, nist-cvss}, administrators may struggle to identify the most critical alerts amid noise or assess their broader impact on infrastructure. For instance, in Figure~\ref{f:use_case_apt}, an administrator may overlook security alerts related to early attack attempts (\circled{1}), even though these events could lead to severe consequences, such as the leakage of confidential information.

\subsection{Assumptions and Threat Model}
\label{s:prob-threat}

\noindent\textbf{Assumptions.} We assume an SDN-enabled enterprise network in which an administrator defines high-level security requirements and specifies which assets must be protected. To satisfy these requirements, the administrator deploys network security mechanisms using either dedicated security devices (e.g., NIDS/network intrusion prevention systems (NIPS)) or SDN controllers, thereby enabling flow-level access control and monitoring. However, owing to the dynamic nature of enterprise networks---such as topology changes or policy conflicts across devices---we assume that misconfigurations or human errors may lead to discrepancies between the intended high-level requirements and their actual enforcement, potentially enabling APTs by adversaries. \\

\noindent\textbf{Threat Model.} We consider adversaries who attempt to gain unauthorized access to internal assets in an enterprise network with malicious intent, such as stealing sensitive information, disrupting services, or escalating privileges, to attack protected resources. These adversaries may employ a range of penetration techniques, including remote exploitations, bypassing security devices, leveraging stepping stones, or compromising network components such as switches and links. To focus on the core objectives of {\ourtool}, we assume that adversaries cannot compromise or manipulate {\ourtool} itself (e.g., its internal logic or execution), nor can they tamper with the trusted components responsible for collecting provenance data. However, we assume that adversaries can influence the network environment (e.g., via ARP spoofing), thereby affecting the accuracy or completeness of the provenance data collected by \ourtool{}.


\section{Related Work}
\label{s:related}
Provenance-based forensic and IDS are increasingly applied across various domains, extending beyond traditional networks to environments such as IoT and cloud computing~\cite{pohly2012hi, gehani2012spade, bates2015trustworthy, pasquier2017practical, chen2021clarion, datta2022alastor, tabiban2022provtalk}. These systems operate by gathering host events, such as system calls~\cite{pasquier2017practical, irshad2021trace, dong2023we, sekar2024eaudit}, and analyzing attack attempts based on the causal relationships between the collected events. This approach facilitates the effective clarification of the relationships between events, thereby enhancing the capability of detecting more intricate attack patterns. These PIDS identify attacks with one of the two primary types: (i) anomaly-based or (ii) rule-based attacks. \\

\begin{table*}
    \centering
    \caption{Comparison of previous provenance-based forensic and IDS and \ourtool{}.\\(\fullcircle \hspace{0.01in} Fully applicable, \halfcircle \hspace{0.01in} Partially applicable, and \emptycircle \hspace{0.01in} Not applicable)}
    
    \resizebox{\linewidth}{!}{
    \scriptsize
    \begin{tabular}{c c c c c c}
    \toprule
         \textbf{Type} & \textbf{Work} & \textbf{Target} & \textbf{SDN-based Design} & \textbf{Network-level Provenance} & \textbf{Automatic RCA} \\ \midrule

         \multirow{8}{*}{Forensic Systems}

          & Hi-Fi~\cite{pohly2012hi} & OS Kernel & \emptycircle & \emptycircle & \emptycircle \\ \cmidrule{2-6}
          
          & SPADE~\cite{gehani2012spade} & OS Kernel & \emptycircle & \emptycircle & \emptycircle \\ \cmidrule{2-6}
          
          & LPM~\cite{bates2015trustworthy} & OS Kernel & \emptycircle & \emptycircle & \emptycircle \\ \cmidrule{2-6}
          
          & CamFlow~\cite{pasquier2017practical} & Cloud & \emptycircle & \halfcircle & \emptycircle \\ \cmidrule{2-6}
          
          & Clarion~\cite{chen2021clarion} & Microservice & \emptycircle & \fullcircle & \emptycircle \\ \cmidrule{2-6}
          
          & Alastor~\cite{datta2022alastor} & Serverless & \emptycircle & \fullcircle & \emptycircle \\ \midrule

          \multirow{18}{*}{Anomaly-based Detection}
          & Poirot~\cite{milajerdi2019poirot} & Enterprise Host & \emptycircle & \emptycircle & \halfcircle \\ \cmidrule{2-6}

          & Nodoze~\cite{hassan2019nodoze} & Enterprise Host & \emptycircle & \emptycircle & \fullcircle \\ \cmidrule{2-6}
          
          & Unicorn~\cite{han2020unicorn} & Enterprise Host & \emptycircle & \emptycircle & \emptycircle \\ \cmidrule{2-6}
          
          & ProvDetector~\cite{wang2020you} & Malware & \emptycircle & \emptycircle & \emptycircle \\ \cmidrule{2-6}
          
          & Atlas~\cite{alsaheel2021atlas} & Enterprise Host & \emptycircle & \halfcircle & \fullcircle \\ \cmidrule{2-6}

          & ThreaTrace~\cite{wang2022threatrace} & Enterprise Host & \emptycircle & \emptycircle & \emptycircle \\ \cmidrule{2-6}
          
          & ShadeWatcher~\cite{zengy2022shadewatcher} & Enterprise Host & \emptycircle & \emptycircle & \emptycircle \\ \cmidrule{2-6}
          
          & Prographer~\cite{yang2023prographer} & Enterprise Host & \emptycircle & \halfcircle & \halfcircle \\ \cmidrule{2-6}
          
          & Magic~\cite{jia2024magic} & Enterprise Host & \emptycircle & \emptycircle & \emptycircle \\ \cmidrule{2-6}

          & R-CAID~\cite{goyal2024r} & Enterprise Host & \emptycircle & \emptycircle & \fullcircle \\ \cmidrule{2-6}
          
          & Kairos~\cite{cheng2024kairos} & Enterprise Host & \emptycircle & \halfcircle & \fullcircle \\ \cmidrule{2-6}
          
          & Flash~\cite{rehman2024flash} & Enterprise Host & \emptycircle & \halfcircle & \fullcircle \\ \midrule

          \multirow{14}{*}{Rule-based Detection}
          & SLEUTH~\cite{hossain2017sleuth} & Enterprise Host & \emptycircle & \halfcircle & \halfcircle \\ \cmidrule{2-6}

          & Holmes~\cite{milajerdi2019holmes} & Enterprise Host & \emptycircle & \halfcircle & \halfcircle \\ \cmidrule{2-6}
          
          & MORSE~\cite{hossain2020combating} & Enterprise Host & \emptycircle & \halfcircle & \halfcircle \\ \cmidrule{2-6}
          
          & RapSheet~\cite{hassan2020tactical} & Enterprise Host & \emptycircle & \halfcircle & \halfcircle \\ \cmidrule{2-6}
          
          & ProvTalk~\cite{tabiban2022provtalk} & NFV & \emptycircle & \halfcircle & \fullcircle \\ \cmidrule{2-6}
          
          & ForenGuard~\cite{wang2018towards} & SDN Attacks & \fullcircle & \halfcircle & \emptycircle \\ \cmidrule{2-6}

          & ProvSDN~\cite{ujcich2018cross} & SDN Attacks & \fullcircle & \halfcircle & \halfcircle \\ \cmidrule{2-6}

          & P4Control~\cite{bajaber2024p4control} & Enterprise Network Intrusions & \emptycircle & \fullcircle & \fullcircle \\ \cmidrule{2-6}

          & \ourtool{} (our work) & Enterprise Network Intrusions & \fullcircle & \fullcircle & \fullcircle \\
    \bottomrule
    \end{tabular}
    }
    
    \label{tab:related_work}
\end{table*}

\noindent\textbf{Anomaly-based.} Anomaly-based provenance systems focus on the construction and analysis of provenance graphs that encapsulate the relationships and interactions between events occurring in the host~\cite{milajerdi2019poirot, hassan2019nodoze, han2020unicorn, wang2020you, alsaheel2021atlas, wang2022threatrace, zengy2022shadewatcher, yang2023prographer, jia2024magic, goyal2024r, cheng2024kairos, rehman2024flash}. Using advanced graph machine learning algorithms, these systems analyze provenance graphs to detect anomalies that signify deviations from established normal behavior, indicating potential intrusions. This approach is particularly effective in identifying zero-day exploits and sophisticated malware that may not match known signatures or rules. \\

\noindent\textbf{Rule-based.} By contrast, rule-based provenance systems employ a methodology in which collected host events are meticulously matched against an established set of rules that define the characteristics of attacks~\cite{hossain2017sleuth, milajerdi2019holmes, hossain2020combating, hassan2020tactical}. These rules model APT attacks as sequences of malicious actions, thereby allowing for the systematic detection of similar attack patterns within the network. Moreover, these systems enrich the detection process by correlating identified attacks with comprehensive attack-related databases such as the MITRE ATT\&CK framework. This correlation not only facilitates a more detailed understanding of attack tactics, techniques, and procedures but also enhances the contextualization of security alerts, making them more actionable for security analysts.

Recent innovations in this domain have integrated network event monitoring to provide holistic security overviews. For example, P4CONTROL~\cite{bajaber2024p4control} is a pioneering tool that leverages eBPF for the collection of host events and extends its surveillance capabilities to observe network events between hosts within programmable switches. By applying rule-based analytics, this tool enhances the ability to detect and respond to complex attacks across multiple hosts in both host and network layers. PivotWall~\cite{oconnor2018pivotwall} proposed a security framework that integrates information flow tracking with SDN to detect and mitigate APTs and stepping-stone attacks within enterprise networks. \\

\noindent
\textbf{Comparison with \ourtool{}.} In response to the evolving landscape of network security, \ourtool{} was specifically designed for enterprise networks, distinguishing itself from traditional systems that primarily focus on single-host events. In contrast to these conventional approaches, \ourtool{} conducts comprehensive network-wide monitoring across multiple hosts using an SDN to facilitate an in-depth RCA. This capability enables a more precise understanding of network dynamics and anomalies, allowing administrators to trace the origins of security incidents accurately. By offering a specialized approach to network event analysis, \ourtool{} significantly enhances threat detection and mitigation through SDN infrastructure, thereby strengthening the overall security posture of these critical networks. Table~\ref{tab:related_work} summarizes our analysis results, which compare \ourtool{} with previous solutions in terms of (i) SDN-based centralized monitoring, (ii) network-level security provenance, and (iii) automatic RCA, which are the main design considerations.


\section{{\ourtool} Design}
\label{s:design}

In this section, we present the high-level approach of \ourtool{} and describe its architecture and key components.

\subsection{High-level Approaches}
\label{s:sp-sp}

\ourtool{} enables administrators to perform {\em{RCA}} to identify potential security issues in enterprise networks by assessing their impact on the infrastructure. To this end, \ourtool{} considers three high-level approaches (A1--A3):\\ 

\noindent\textbf{A1. SDN-based Centralized Monitoring.}
{\ourtool} aims to achieve information-lossless forensics to ensure that no security events or issues are overlooked, thus enabling administrators to accurately reconstruct the events occured during a security incident. To achieve this, \ourtool{} leverages the centralized visibility and flow-level monitoring capabilities of SDN~\cite{lee2017athena,shin2016enhancing,kim2024enhancing}, enabling consistent tracking of host-to-host communication across complex enterprise environments. Although SDN traffic data provide broad coverage, they are insufficient to fully explain the cause, method, and location of security incidents. To address this, \ourtool{} collects a variety of \textit{security snapshots}, including security policies, network configurations, host profiles, network topology (including host locations), forwarding rules, and provenance monitoring data captured at specific points in time. These datasets, collectively termed as \textit{security provenance data}, serve as the foundation for security diagnosis, enabling the identification of security causalities and providing a comprehensive view of attack progression. \\

\noindent\textbf{A2. Network-wide Provenance Graph Construction.}
\ourtool{} introduces a \emph{network-level security provenance graph} that captures interhost security events in an enterprise network. Traditional host-level causality delineates the causal relationships among events associated with the host system resources, including processes, files, and network sockets. We generalize this concept to the network level, enabling the representation of causal dependencies between network security events across diverse hosts within an enterprise environment, such as servers, databases, employee workstations, and administrative systems.
Consequently, an entire attack sequence over a network is represented as a series of historical security causalities, illustrating how each host is affected and how the attack progresses over time. For example, in Figure~\ref{f:use_case_apt}, the red arrows represent causalities, providing administrators with critical insights into the flow of an attack and enabling them to recognize potential information leaks. \\

\noindent\textbf{A3. Automatic RCA.}  
{\ourtool} prioritizes endangered network assets in enterprise networks through \emph{rule-based root cause analyses}, thereby enabling administrators to quickly identify and inspect the most critical assets in response to security incidents. To achieve this, we leveraged \textit{PSL}, a framework that is widely used in areas such as natural language processing, social network analysis, and knowledge graphs~\cite{kimmig2012short}. Our approach organizes various network and security events into a relational form and applies PSL to determine which hosts are attacked, which are the potential targets, and what types of high-risk incidents have occurred. Prioritization is guided by predefined rules to ensure an efficient incident response. For example, in Figure~\ref{f:use_case_apt}, the internal host controlled by the attacker is classified as both a \textit{target} and a \textit{suspect}, which helps administrators quickly identify critical assets at risk. This approach differs significantly from traditional solutions, which primarily assess security events in isolation based on their severity scores~\cite{snort-class, nist-cvss} without considering their broader context or relational impact.

\subsection{{\ourtool} Overview}
\label{s:design-overview}
\begin{figure}
\centering
   \includegraphics[width=1\linewidth]{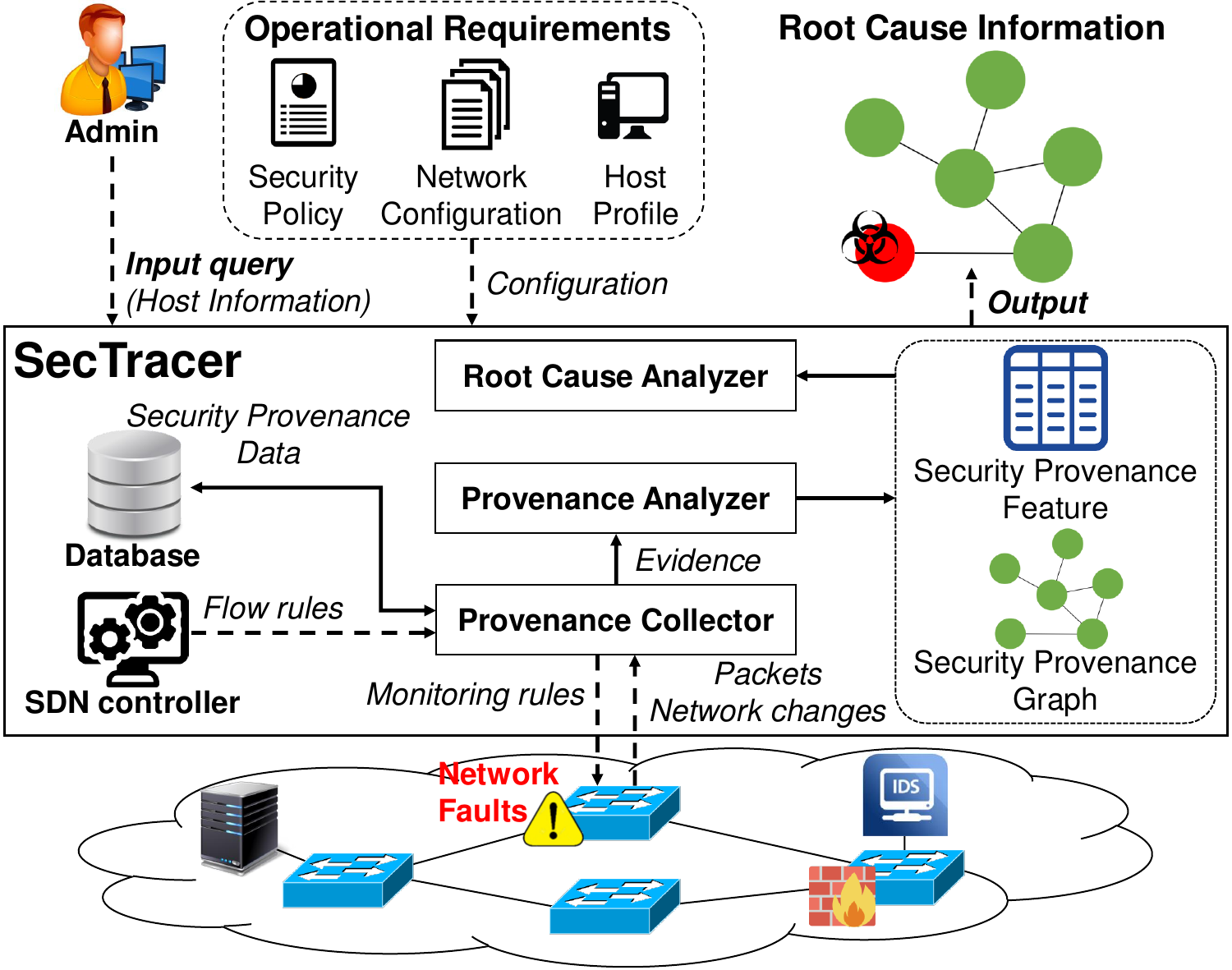}
    \caption{{Overall architecture of \ourtool}. The {\ourtool} is hosted over a wide enterprise infrastructure above the SDN control plane. The solid and dotted lines indicate an internal and external event, respectively.}
    \label{f:overall_architecture}
    \vspace{-0.1in}
\end{figure}

Figure~\ref{f:overall_architecture} illustrates the architecture of {\ourtool}, which was designed based on the proposed approach. At a high level, an administrator must provide \ourtool{} with three key operational requirements: (i) {\em{security policy}}, which defines a routing policy specifying which hosts should pass through specific security devices and be subject to access control rules, (ii) {\em{network configurations}}, which include details about the locations of security devices or middleboxes within the network, and (iii) {\em{host profile}}, which specifies which security provenance data should be monitored for certain hosts and outlines the security policies governing them. Based on these inputs, \ourtool{} constructs a {\em{security provenance graph}} enriched with root cause information to aid the security analysis.

{\ourtool} comprises three main modules: (i) a \emph{provenance collector}, (ii) a \emph{provenance analyzer}, and (iii) a \emph{root cause analyzer}. The \emph{provenance collector} gathers the security provenance data, generates supporting evidence, and publishes both to a database. For provenance monitoring, {\ourtool} leverages SDN to efficiently collect network-related information, such as topology and packet traces. With SDN-centralized network management, the provenance collector tracks network changes by monitoring SDN control messages and issues monitoring flow rules to capture packets corresponding to installed or removed flow rules using the SDN control plane\footnote{We used OpenFlow v1.3 as the southbound SDN protocol~\cite{openflow-1-3}.}. The \emph{provenance analyzer} receives relevant evidence from the provenance collector based on the input host information and constructs a security provenance graph by identifying security causalities among the collected provenance data. Finally, the \emph{root cause analyzer} processes the provenance graph to pinpoint the root cause of security incidents.

Next, Section~\ref{s:design-collector}, Section~\ref{s:design-analyzer}, and Section~\ref{s:design-rca} present the details of {\ourtool} for the provenance collector,provenance analyzer, and root cause analyzer, respectively.


\section{Provenance Collection}
\label{s:design-collector}

The main purpose of the \textit{provenance collector} is to collect security provenance data across an enterprise infrastructure and create evidence with the corresponding security snapshots. This approach incorporates three major components: \textit{packet monitoring}, \textit{security snapshot recording}, and \textit{evidence generation}.

\subsection{Packet Monitoring}
\label{s:design-collector-monitor}

The provenance collector reads the configurations specified in the host profile provided by the administrator, detects flow-rule events from network applications running on the SDN controller, and issues provenance-monitoring rules to capture raw packets in the data plane. \\

\noindent\textbf{Two-Phase Packet Capture.} To enhance packet tracking, we proposed a novel method called {\em{two-phase packet capture}} ({\tpc}) that is designed to monitor both the integrity and reachability of packets. This approach enables monitoring at two distinct strategic points to maximize packet visibility. It operates by duplicating packets at the egress and ingress switches using monitoring rules and forwarding both copies to the provenance collector for further analysis. 

\begin{figure}
    \centerline{\includegraphics[width=1\linewidth]{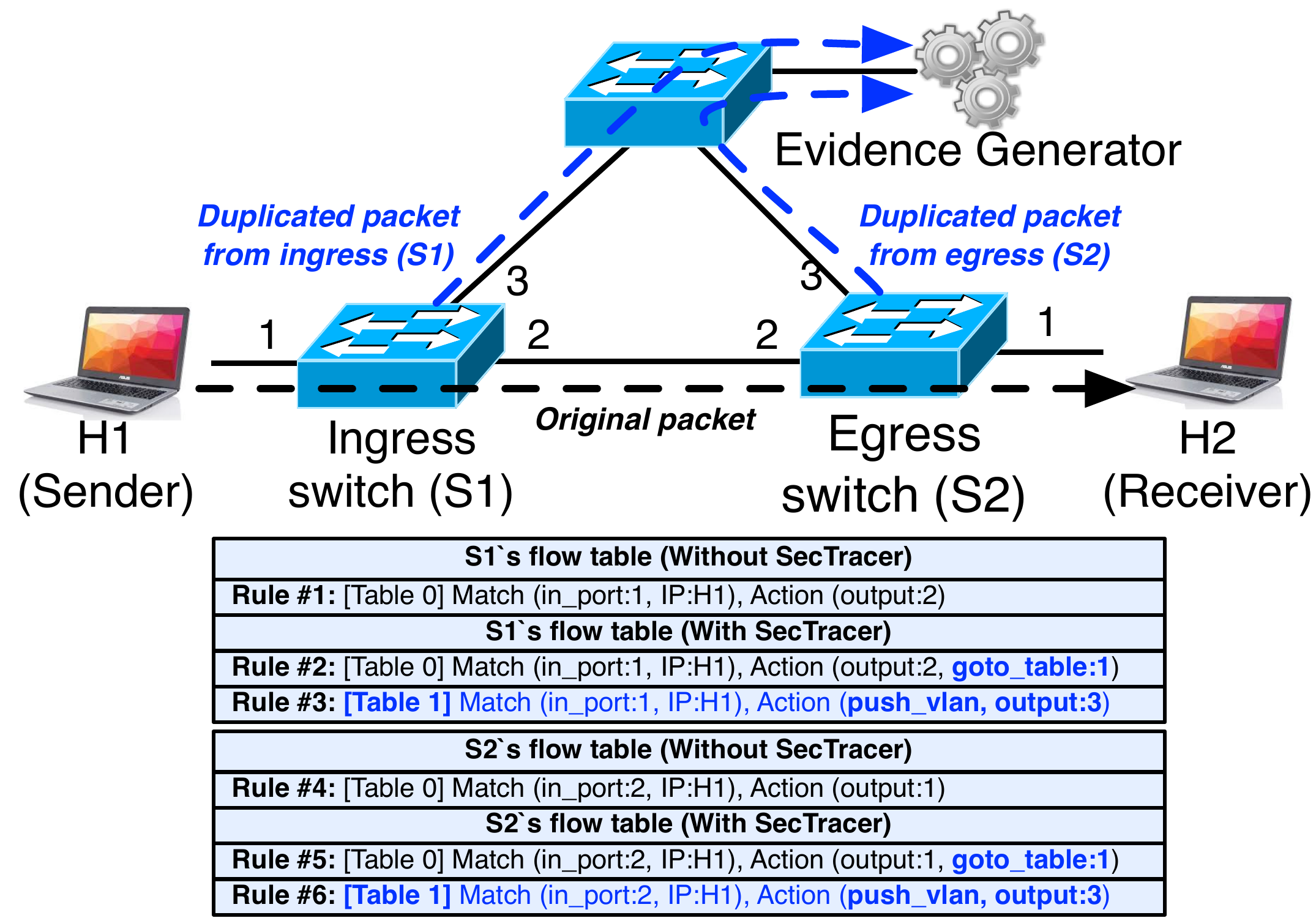}}
    \caption{Working scenario of {\tpc} when \textit{H1} sends a packet to \textit{H2}.  Monitoring parts are highlighted in blue.} 
    \label{f:two_pc}
\end{figure}

Figure~\ref{f:two_pc} illustrates how {\ourtool} leverages SDN techniques, specifically OpenFlow commands\footnote{We use the \texttt{flow\_mod} action to install the flow rules.} to capture packets at both the ingress and egress points of the flow. In this scenario, \textit{H1} sends a packet to \textit{H2}, and switches \textit{S1} and \textit{S2} have flow rules in place to forward the packet from \textit{H1} to \textit{H2} (Rules \#1 and \#4). The packet monitor is assumed to have preinstalled monitoring rules for {\tpc} using OpenFlow commands (Rules \#2, \#3, \#5, and \#6). When a packet arrives at the ingress switch, \textit{Rule \#2} duplicates it using two actions: \texttt{output} and \texttt{goto\_table}. Subsequently, \textit{Rule \#3} applies the \texttt{push\_vlan} action to tag a duplicated packet and forwards it to the evidence generator over an isolated network. Similarly, when the egress switch (\textit{S2}) receives a packet forwarded by \textit{S1}, its monitoring rules operate in the same manner as those of the ingress switch, ensuring comprehensive packet tracking across the flow path. \\

\noindent\textbf{Deploying Monitoring Rules.}
To deploy monitoring rules with {\tpc}, the provenance collector maintains a list of hosts based on the host profile specified in the operational requirements\footnote{Each host profile is defined using CIDR notation.}. Whenever a network application or administrator issues flow rules\footnote{For flow rules installed or removed by network applications, {\ourtool} monitors the \textit{flow\_mod} and \textit{flow\_removed} messages.}, the packet monitor validates whether the IPv4 address ranges specified in the host profile match those specified in the issued flow rules. If a match is found, the current topology information is retrieved to identify the ingress and egress switches of the host pairs, as well as the edge switch connected to the evidence generator. To locate these switches, {\ourtool} leverages a network-invariant checker to determine the end-to-end paths~\cite{khurshid2013veriflow,kazemian2012header} and considers packet detouring through network middleboxes, as specified in the network configuration, within the operational requirements\footnote{The network configuration defines the locations of middleboxes.}. When the relevant switches are identified, the provenance collector issues monitoring rules to enable packet duplication and forwarding, thereby ensuring comprehensive security provenance tracking.

\subsection{Security Snapshot Recording}
\label{s:design-collector-snap}
Changes in the operational requirements generate \textit{configuration change} events that notify the provenance collector of updates for security snapshots. The network changes track the topology modifications (e.g., switch and port up/down), host location changes, and flow table updates for each switch. Because popular SDN controllers, such as ONOS~\cite{berde2014onos} and OpenDaylight~\cite{opendaylight}, maintain local states for topology information, host locations, and internal flow tables per switch, the provenance collector reads these local states to monitor and track network changes effectively. 
The provenance collector maintains historical security snapshots to maintain the internal states while monitoring the packets. Technically, whenever any security policies, network configurations, host profiles, topology including host locations, flow entries in each switch, and packet monitoring information are changed at a particular point in time, provenance collector publishes an updated snapshot of {\ourtool}'s database with a unique identifier specifying its version.

\begin{figure}
    \centering
       \includegraphics[width=1\linewidth]{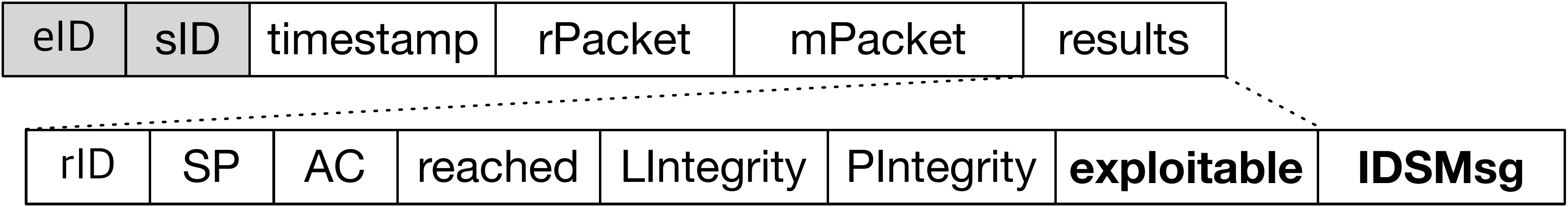}
        \caption{Format of a single {\ourtool} evidence. The gray box represents index fields for unique identifiers, and the white box is a list of feature fields. The text in bold represents features from the security devices, and the others in the white box are generated from enterprise-wide security inspections.}
    \label{f:evidence}
\end{figure}

\subsection{Evidence Generation}
\label{s:design-collector-evidence}

The provenance collector generates evidence by analyzing a monitored packet and its corresponding security snapshot. Figure~\ref{f:evidence} illustrates the evidence format used by {\ourtool}. Upon receiving two mirrored packets\footnote{If a packet is dropped, a single copy is forwarded to \ourtool{} by the ingress switch.}, \ourtool{} assigns unique identifiers, \emph{eID} and \emph{sID}, to the evidence and the associated security snapshot, respectively. It records the monitoring time in the \emph{timestamp} field and then performs a security diagnosis on the provenance data to populate the feature fields. To ensure high-quality security diagnosis results, that is, accurate feature extraction, {\ourtool} employs two key strategies: (i) {\em{leveraging security devices}} and (ii) {\em{network-wide security inspections}}. Details of these strategies are discussed as follows. \\

\noindent\textbf{Leveraging Security Devices.}  
To ensure compatibility with enterprise networks in which security devices have already been deployed, \ourtool{} integrates existing security mechanisms, including network intrusion detection and prevention systems (NIDS/NIPS) and anomaly detection systems, as provenance sources. Specifically, the provenance collector leverages a software-based NIDS~\cite{roesch1999snort, suricata} by forwarding monitored packets (i.e., packet streams) to an IDS instance for analysis. When an alert is triggered, the provenance collector assesses whether the detected activity corresponds to an exploitable attack\footnote{We used Snort \emph{classtype} to assess an attack's exploitability. If the \emph{classtype} of the Snort alert is \emph{high}, {\ourtool} classifies the attack as exploitable.}. If the attack is deemed exploitable, the \emph{exploitable} field is set to \emph{true}; otherwise, it is set to \emph{false}. Finally, the raw alert messages are stored in \emph{IDSMsg}, providing users with detailed insights into the detected threats. \\

\noindent\textbf{Network-wide Security Inspections.}
Existing security devices have limited coverage because of their strategic deployment at network choke points, which restricts them from inspecting only a subset of the network flows~\cite{shin2012cloudwatcher,fayaz2015bohatei,lee2017athena}. Even when administrators have full control over individual host machines, they lack visibility of network-wide activities. To enable comprehensive network-wide monitoring, {\ourtool} evaluates two key security aspects. 

First, it monitors the {\em reachability} to determine whether an attacker’s attempt successfully traverses the network under the current configuration. In particular, reachability is assessed by identifying a pair of packets that originate from the same source and comparing the immutable fields in the monitored packets~\cite{handigol2014know}. By leveraging {\tpc}, a provenance collector receives one or two packets per attempt. If a matching pair is observed, the \emph{reached} field is set to \emph{true}. Otherwise, if no corresponding packet is found, the \emph{reached} field is set to \emph{false}, which indicates that the packet is dropped in the network.

Second, it checks for {\em policy violations} to ensure that the administrator’s high-level security requirements are correctly enforced without errors or misconfigurations. To achieve this, \ourtool{} examines whether any monitored packet violates the security policies defined in the operational specifications. Because the monitoring data include a list of middleboxes traversed by each packet, \ourtool{} compares the observed path with the intended policy to verify the compliance. It also verifies whether a packet has been delivered to unauthorized hosts by validating it against the access control policy. If a violation is found in either the security routing policy or the access control policy, the fields \texttt{SP} and \texttt{AC} are set as the identifiers of the corresponding violated policies\footnote{Each policy entry is assigned a unique identifier.}.\\

\begin{figure}
    \centerline{\includegraphics[width=\linewidth]{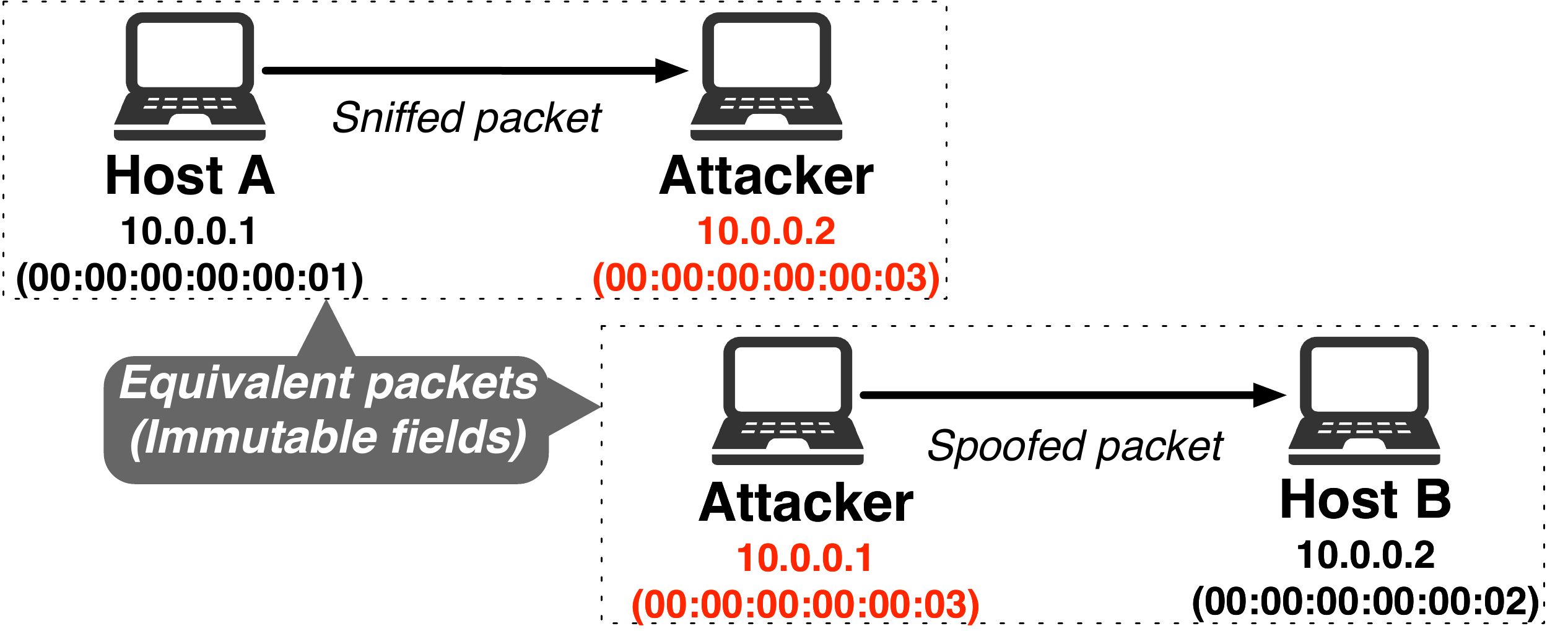}}
    \caption{Example scenario of a local integrity violation between Hosts A and B, where an attacker intercepts a packet from Host A, spoofs its IP address to appear as Host A, and sends it to Host B.}
    \label{f:logical_integrity_check}
\end{figure}

\noindent
\textbf{Tamper-Resistant Provenance Collection.} When \ourtool{} collects provenance data, an attacker may attempt to tamper with it by manipulating packet headers such as IP addresses (Section~\ref{s:prob-threat}). To ensure tamper-resistant provenance collection, \ourtool{} detects integrity violations categorized into {\em physical} and {\em logical} types. A {\em physical integrity violation} occurs when an attacker manipulates a packet by compromising physical network components, such as links or switches. A packet is considered physically violated if (i) it is not dropped, and (ii) the monitored versions of the packet contain differing headers or payloads; in such cases, the \emph{PIntegrity} flag is set to \emph{true}. In contrast, a {\em logical integrity violation} involves intercepting and relaying packets between hosts, as in man-in-the-middle attacks. Figure~\ref{f:logical_integrity_check} illustrates an example scenario\footnote{\ourtool{} currently supports ARP spoofing as a concrete example of logical integrity violations. Other MITM-style attacks, such as DNS and IP spoofing, follow similar patterns and can be supported.} in which an attacker intercepts the packets, modifies the header fields, and replays them. A logical integrity violation is detected if (i) two packets within a connection share the same immutable fields, and (ii) a single MAC address is associated with two different IP addresses. When both conditions are satisfied, \emph{LIntegrity} is set to \emph{true}, and the attacker’s identifier (\emph{eID}) is stored as a reference ID (\emph{rID}) in the corresponding evidence record. For both violation types, the provenance collector stores the original and manipulated packets in the \emph{rPacket} and \emph{mPacket} fields, respectively.


\section{Provenance Analysis}
\label{s:design-analyzer}

The \emph{provenance analyzer} constructs a provenance graph that captures security causalities based on the collected evidence and generates security provenance features.

\begin{figure}
\centering
   \includegraphics[width=1\linewidth]{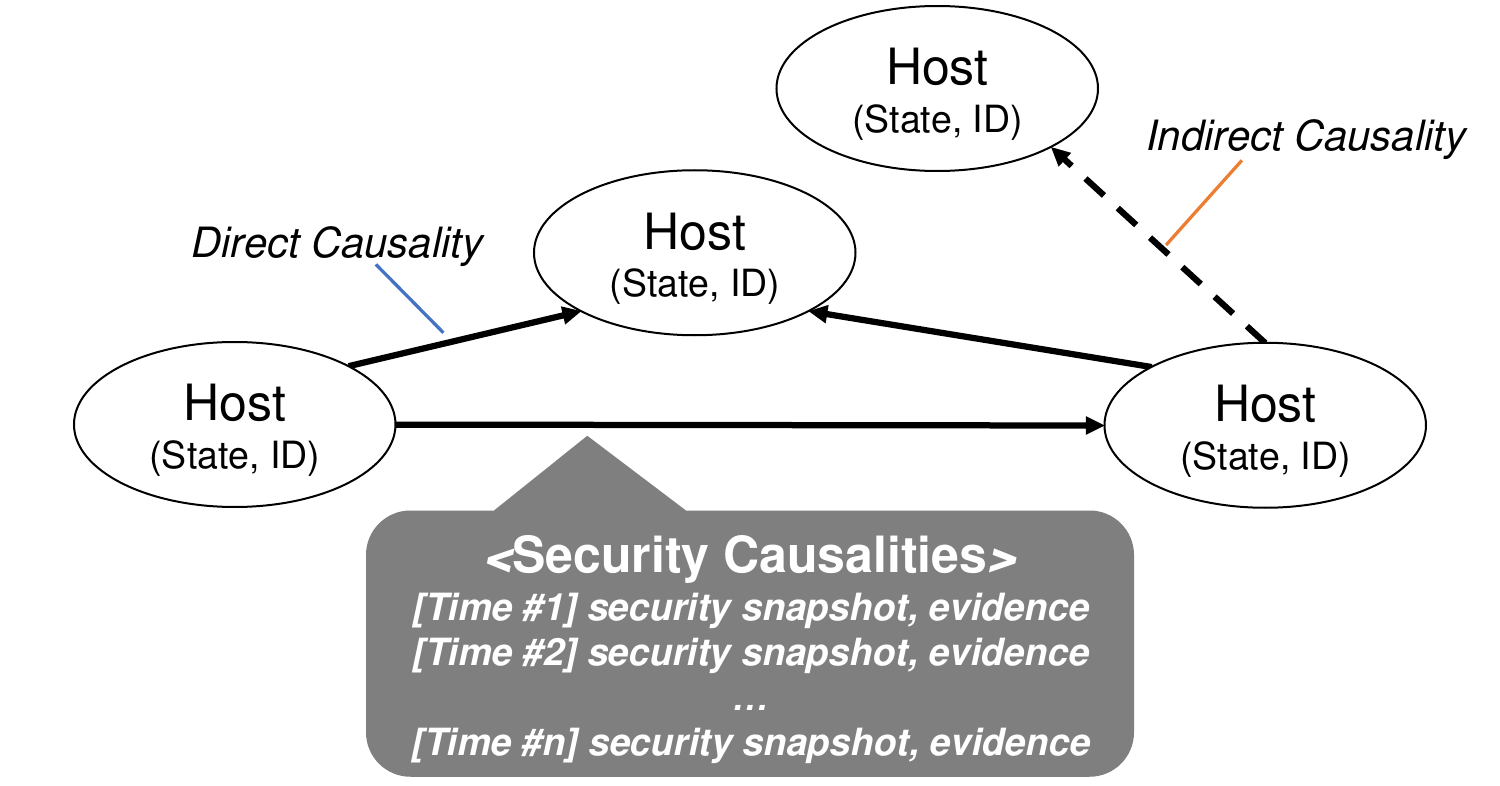}
    \caption{Example security provenance graph, which represents a series of security causalities, from a given host as an input. A vertex indicates a host with the unique identifier and the state. An edge represents security causalities between the hosts with corresponding evidence(s) and their security snapshot(s) in chronological order. A solid line represents a directed causality, and a dotted line represents an indirect causality.}
    \label{f:provenance_graph_overview}
\end{figure}

\subsection{Security Provenance Graph}
\label{s:design-analyzer-graph}

The provenance analyzer constructs a \emph{network-level security provenance graph}, which is a specialized data structure that captures the complete sequence of attack procedures and provides valuable insights into the causes of security incidents across hosts. By analyzing this graph, administrators can effectively understand past security-related events without an extensive manual investigation. It offers a comprehensive view of host activities and operational behaviors, enabling administrators to analyze how an attack was executed and identify its root cause.

Figure~\ref{f:provenance_graph_overview} shows an example of a security provenance graph. Each vertex represents a host that serves as either a subject or an object within the graph. A vertex has two properties: (i) {\em{state}}, which indicates whether the host is a \textit{victim} or an \textit{attacker}, and (ii) \textit{identification (ID)}, which consists of an {\em{IP address}} and a {\em{MAC address}}, serving as a unique identifier for the host.
An edge represents a \emph{security causality} between two hosts and contains security snapshot properties that store system snapshots at a specific time when the corresponding security causality is established. 
Multiple instances of security causality can occur between the same pair of nodes, resulting in multiple edges in the provenance graph. As each edge embeds temporal information, it can be leveraged for detailed attack reconstruction and forensic analysis.

The provenance analyzer constructs a provenance graph based on the IP address of the host, as specified by the administrator. It retrieves the evidence records in which the packet’s IP address corresponds to a given input.
Based on the retrieved evidence, it establishes security causalities as edges, connects the corresponding subject and object hosts as vertices, and assigns states to each host according to the defined security causalities (Section~\ref{s:design-graph-causality}). After updating the security provenance graph, the provenance analyzer selects the next attack source and iteratively expands the graph until no further targets remain.

\subsection{Security Causality}
\label{s:design-graph-causality}

\begin{table}[t]
\scriptsize
\caption{Enumeration of supported security causalities. Star-notation(\textsuperscript{*}) indicates \textit{indirect causalities}.}
\begin{center}
\begin{tabular}{l|l}
\bf{Security Causality} & \multicolumn{1}{c}{\bf{Description}}\\
\hline 
\hline 
\#1 connect\_to & Connection to arbitrary hosts\\ \hline 
\#2 attack\_to & Exploitable attack detected\\ \hline 
\#3 attack\_trial\_to & Attack trial detected\\ \hline 
\#4 attack\_blocked & Attack (trial) blocked\\ \hline 
\#5 p\_insecure\_connect\_to & Physical integrity violation\\ \hline 
\#6 l\_insecure\_connect\_to & Logical integrity violation\\ \hline 
\#7 violate\_integrity\_of & Subject of integrity violation\\ \hline 
\#8 violate\_ac & Access control policy violation\\ \hline 
\#9 violate\_sp & Security policy violation\\ \hline 
\#10 risky\_connect\_to\textsuperscript{*} & Connection to an identified attacker\\ \hline 
\#11 implicit\_violate\_ac\textsuperscript{*} & Potential access control policy violation\\ \hline 
\#12 implicit\_violate\_sp\textsuperscript{*} & Potential security policy violation\\ 
\end{tabular}
\end{center}

\label{t:security_causality}
\end{table}

\noindent\textbf{Direct and Indirect Causality.} Security causality can be categorized into two types: direct and indirect. {\em{Direct causality}} is established when strong evidence links an action to its consequences. For example, the attack trials in Figure~\ref{f:use_case_apt} illustrate the direct causality because they clearly depict the step-by-step progression of an attack. In contrast, {\em{indirect causality}} lacks definitive evidences to deterministically establish causality but suggests a potential impact on the target hosts. For instance, in our motivating example, the attacker has indirect causality with the database or web server because access is gained through the manager~\cite{zhang2000detecting}. In total, {\ourtool} supports 12 types of security causalities, as listed in Table~\ref{t:security_causality}.

\begin{figure}
\centering
   \includegraphics[width=.9\linewidth]{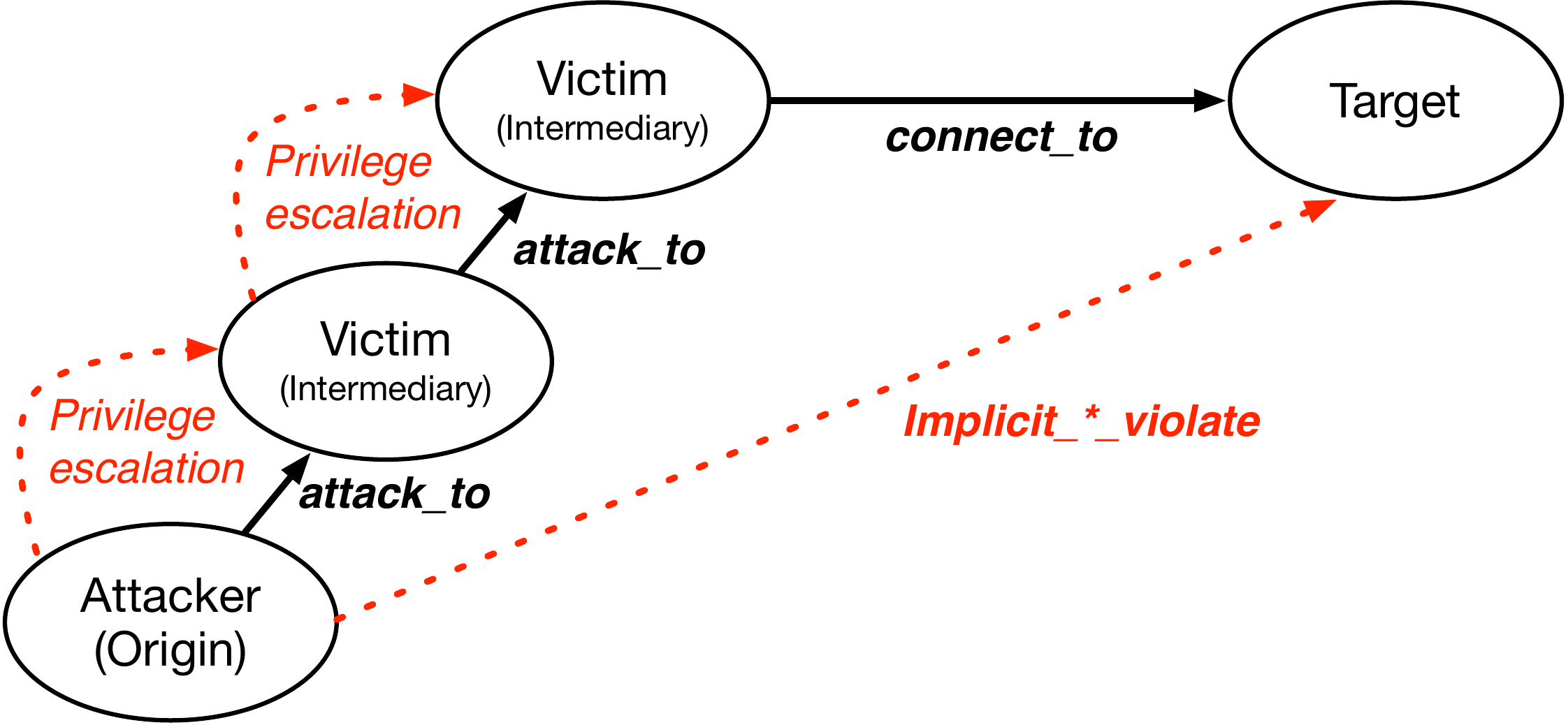}
    \caption{Example of security causality between attacker and target. The dotted line indicates an indirect security causality over intermediary victims.}
    \label{f:implicit_security_causality}
\end{figure}

\begin{figure}
\centering
   \includegraphics[width=.9\linewidth]{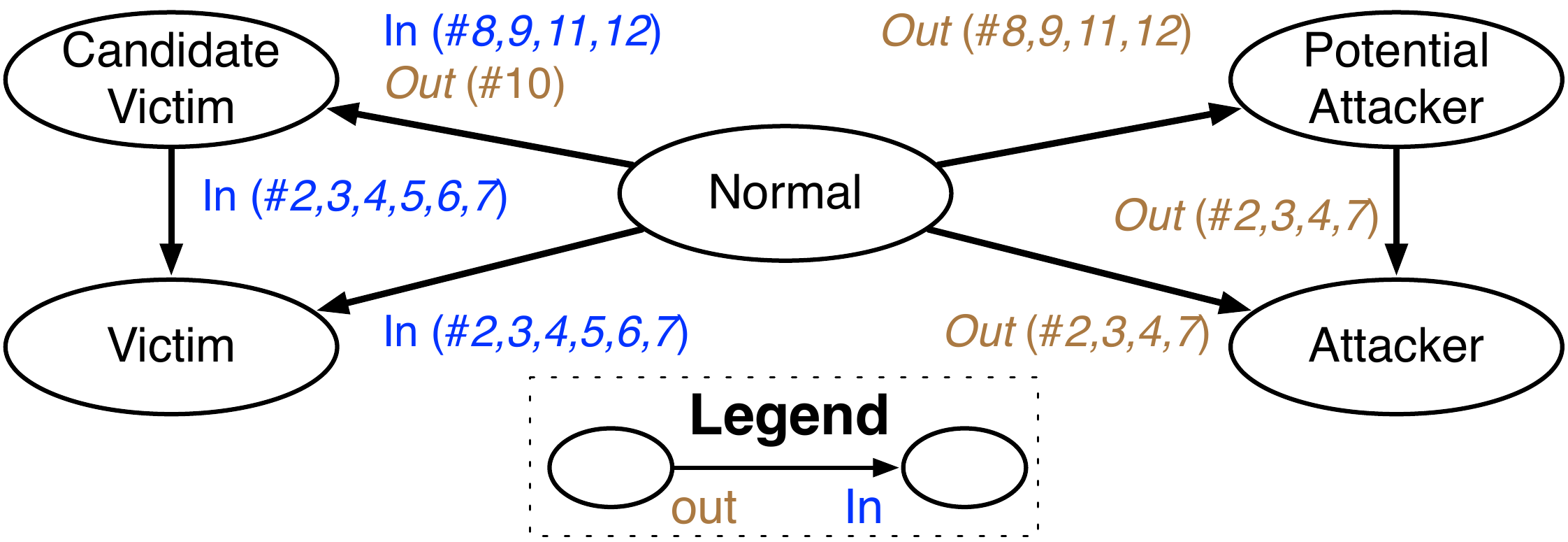}
    \caption{Conditions of state transition. Each number indicates the unique number of each security causality in Table~\ref{t:security_causality}.}
    \label{f:state_transition}
\end{figure}

Figure~\ref{f:implicit_security_causality} illustrates the indirect causality involving intermediary victims whose privileges have been compromised by attackers. Although \emph{attack\_to} does not provide definitive evidence that victims have been fully compromised, the provenance analyzer flags these cases as potential risks, indicating that an attacker has attempted to penetrate the victim’s system. The provenance analyzer follows a chain of \emph{attack\_to} relationships to identify all intermediary victims and considers the leaf node of the chain to have potential access to the victim's associated hosts. If a connection between the leaf node and the victim’s associated hosts violates security policies, the provenance analyzer generates \emph{implicit\_ac\_violate} and \emph{implicit\_sp\_violate} to indicate implicit access control and security policy violations, respectively. In addition, to warn about risky connections that may lead to further security breaches, it generates \emph{risky\_connect\_to}, which signifies that a host has established a connection with a known attacker who has already triggered security causalities, leading to a state transition to the \emph{Attacker }. \\

\noindent\textbf{Vertex State Transition.}
Each \emph{security causality} dynamically updates the vertex state as the security provenance graph evolves. As shown in Figure~\ref{f:state_transition}, the \emph{state} of a node is determined by identified causality relationships. When security causality is detected from one node to another, the state of the affected node changes accordingly. The \emph{Attacker} state indicates that a vertex is actively involved in malicious activities, such as initiating attack trails or causing integrity violations. The \emph{Potential Attacker} state indicates that a vertex possesses the capability to violate security policies, although no explicit malicious behavior has yet been observed. The \emph{Candidate Victim} state represents a host that is either accessible to a \emph{(potential) attacker } without proper authorization or has established potentially vulnerable connections with an \emph{ attacker}. The \emph{Victim} state is the most critical because it signifies the definitive evidence linking the host to a confirmed attack trial. In a security provenance graph, each node maintains two independent states---one for its role as an \emph{ attacker } and another for its role as a \emph{ victim }---because a host can act concurrently in both capacities. When a node assumes both the \emph{Attacker} and \emph{Victim} states, this indicates that the node has been exploited as a \emph{stepping stone} that provides valuable context during the subsequent RCA phase.

\subsection{Security Provenance Features}
\label{s:design-analyzer-feature}

\begin{table}[t]
\scriptsize
\caption{\textit{Security Provenance Feature.} Each feature has one or two variables, which represent the network assets (\textit{H}: hosts and \textit{S}: switches).}
\begin{center}
\begin{tabular}{c|l|c|c}
\textbf{Category} & \textbf{SP Feature} & \textbf{Numeric} & \textbf{Boolean} \\
\hline
\hline 
\multirow{4}{5em}{\centering{Connection}}
 & Connection (H1, H2)          &  & \checkmark \\
 & Blocked (H1, H2)             &  & \checkmark \\ 
 & Number of packets (H1, H2)   & \checkmark &  \\
 & Packet bytes (H1, H2)        & \checkmark &  \\
\hline 
\multirow{6}{5em}{\centering{Session}}
 & Sessions (H1, H2)            & \checkmark &  \\
 & Flag SYN (H1, H2)            & \checkmark &  \\ 
 & Flag FIN (H1, H2)            & \checkmark &  \\
 & Session failure (H1, H2)     & \checkmark &  \\
 & Number of packets (H1, H2)   & \checkmark &  \\
 & Packet bytes (H1, H2)        & \checkmark &  \\
\hline 
\multirow{3}{5em}{\centering{Security Device}}
 & IDS result (H1, H2)          & \checkmark &  \\
 & Firewall drop (H1, H2)       &  & \checkmark \\ 
 & IPS drop (H1, H2)            &  & \checkmark \\
\hline 
\multirow{5}{5em}{\centering{Network Integrity}}
 & Physical spoofing (H1, H2)   &  & \checkmark \\
 & Logical spoofing (H1, H2)    &  & \checkmark \\ 
 & Switch connection (H, S)     &  & \checkmark \\
 & Switch failure (S)           &  & \checkmark \\
 & Link failure (S1, S2)        &  & \checkmark \\
\hline 
\multirow{6}{5em}{\centering{Policy}}
 & AC violation (H1, H2)        &  & \checkmark \\
 & SP violation (H1, H2)        &  & \checkmark \\ 
 & Blacklist IP (H1, H2)        &  & \checkmark \\
 & Blacklist protocol (H1, H2)  &  & \checkmark \\
 & Blacklist port (H1, H2)      &  & \checkmark \\ 
 & Blacklist service (H1, H2)   &  & \checkmark \\
\hline 
\multirow{2}{5em}{\centering{Host}}
 & Internal (H)                 &  & \checkmark \\ 
 & Victim (H)                   &  & \checkmark \\
\hline
\end{tabular}
\end{center}
\label{t:sp_feature}
\end{table}

The provenance analyzer generates \textit{security provenance features (SP features)} to support RCA (Section~\ref{s:design-rca}). These features capture various network behaviors, including packet transmissions between hosts, session details, IDS alert messages related to host communications, and the overall network environment status. After constructing the security provenance graph, the provenance analyzer retrieves relevant evidence from the graph and generates the SP features. These features are created on demand when an administrator initiates an RCA for a specific host, thereby minimizing the overhead. Table~\ref{t:sp_feature} lists the SP features currently supported by \ourtool{}, which are categorized into six groups. Additionally, administrators can extend \ourtool{} by defining new SP features to enhance RCA capabilities.

\begin{itemize}  
    \item \textit{Connection}: It represents packet transmissions between hosts. This category includes network behaviors such as the success or failure of packet transmissions, packet size, and the number of packets exchanged between two hosts.  
    \item \textit{Session}: It describes TCP sessions between hosts. It includes session-related information such as the number of sessions, session failures, average number of packets per session, and average packet size.  
    \item \textit{Security Device}: It captures alert messages generated by security devices. These messages contain classifications and priority levels, which are converted into SP features. 
    \item \textit{Network Integrity}: It reflects changes in network conditions, such as switch or link failures and host migrations between switches. This category also includes packet integrity issues, such as spoofing or packet modifications.  
    \item \textit{Policy}: It represents network policies configured by administrators. It includes security policies, access control policies, and basic configurations such as allowed ports and protocols.  
    \item \textit{Host}: It describes host characteristics. The \textit{internal} feature identifies hosts belonging to the internal network, whereas the \textit{victim} feature denotes hosts that have been attacked or are suspected of being targeted.  
\end{itemize}

SP features are represented as soft truth values within the interval \([0,1]\) and are classified into two data types: \textit{numeric} and \textit{ Boolean }. The numeric type can assume any value between 0 and 1, whereas the Boolean type is restricted to 0 or 1. The choice of datatype depends on the nature of the feature; the Boolean type indicates the presence or absence of a specific behavior, whereas the numeric type quantifies its magnitude or frequency. When the numerical values exceed the \([0,1]\) range, an appropriate normalization method is applied to scale them. To ensure that numeric features fell within this range, we computed the average value of the feature in constant time and normalized it accordingly. The transformation follows a predefined normalization equation to map the computed average to a value between 0 and 1. The normalized value of the feature \( v \) can be obtained using 
\begin{equation} 
\label{e:eq1}
v=1-e^{-ax},
\end{equation}
where \( x \) represents the average value of the feature, and \( a \) is a constant parameter that controls the rate of normalization.  


\section{Root Cause Analysis}
\label{s:design-rca}

\begin{table*}
\centering
\scriptsize
\caption{RCA rules categorized by the types of malicious network behaviors that can be detected when the \textit{root cause analyzer} applies these rules. The \textit{weight} of each rule is determined empirically.}
\begin{center}
\begin{tabular}{c|c|l|p{16em} }
\textbf{Category} & \textbf{Weight} & \textbf{Rules} & \textbf{Results} \\
\hline
\hline 
\multirow{3}{4em}{\centering{Basic Rules}}
 & 1 & Victim(A) & Target(A) \\
 & 1 & Connect(A, B) \& Victim(B) & Suspect(A), Target(A) \\ 
 & 1 & Connect(A, B) \& Suspect(A) \& Internal(B) & Target(B) \\
\hline
\multirow{5}{4em}{\centering{Security Device Rules}}
 & 20 & Connect(A, B) \& IDSResult(A, B) \& !Internal(A) & Suspect(A), Target(B) \\ 
 & 20 & Connect(A, B) \& IDSResult(A, B) \& Internal(A) & Suspect(A), Target(B) \\ 
 & 15 & Blocked(A, B) \& IDSResult(A, B) & Suspect(A), Target(A) \\ 
 & 10 & Blocked(A, B) \& FirewallDrop(A, B) & Target(B) \\
 & 10 & Blocked(A, B) \& IPSDrop(A, B) & Target(B) \\
\hline
\multirow{4}{4em}{\centering{Scanning}}
 & 5 & Connect(A, B) \& FlagSYN(A, B) \& SessionFailure(A, B) & Suspect(A) \\ 
 & 5 & Connect(A, B) \& FlagSYN(A, B) \& BlacklistPort(A, B) & Suspect(A) \\ 
 & 5 & Connect(A, B) \& FlagFIN(A, B) \& SessionFailure(A, B) & Suspect(A) \\ 
 & 5 & Connect(A, B) \& FlagFIN(A, B) \& BlacklistPort(A, B) & Suspect(A) \\ 
\hline
\multirow{2}{4em}{\centering{Policy Violation}}
 & 10 & Connect(A, B) \& ACViolation(A, B) & Suspect(A), Target(B) \\ 
 & 10 & Connect(A, B) \& SPViolation(A, B) & Suspect(A), Target(B) \\ 
\hline
\multirow{4}{4em}{\centering{Blacklist (external)}}
 & 5 & Connect(A, B) \& BlacklistIP(A) & Suspect(A), Target(B) \\ 
 & 5 & Connect(A, B) \& BlacklistPort(A, B) \& !Internal(A) & Suspect(A), Target(B) \\ 
 & 5 & Connect(A, B) \& BlacklistProtocol(A, B) \& !Internal(A) & Suspect(A), Target(B) \\ 
 & 5 & Connect(A, B) \& BlacklistService(A, B) \& !Internal(A) & Suspect(A), Target(B) \\ 
\hline
\multirow{3}{4em}{\centering{Blacklist (internal)}}
 & 2 & Connect(A, B) \& BlacklistPort(A, B) \& Internal(A) & Suspect(A), Target(B) \\ 
 & 2 & Connect(A, B) \& BlacklistProtocol(A, B) \& Internal(A) & Suspect(A), Target(B) \\ 
 & 2 & Connect(A, B) \& BlacklistService(A, B) \& Internal(A) & Suspect(A), Target(B) \\
\hline
\multirow{7}{4em}{\centering{Blacklist (blocked)}}
 & 2 & Blocked(A, B) \& BlacklistIP(A) & Suspect(A) \\ 
 & 2 & Blocked(A, B) \& BlacklistPort(A, B) \& !Internal(A) & Suspect(A) \\ 
 & 2 & Blocked(A, B) \& BlacklistProtocol(A, B) \& !Internal(A) & Suspect(A) \\ 
 & 2 & Blocked(A, B) \& BlacklistService(A, B) \& !Internal(A) & Suspect(A) \\ 
 & 2 & Blocked(A, B) \& BlacklistPort(A, B) \& Internal(A) & Suspect(A) \\ 
 & 2 & Blocked(A, B) \& BlacklistProtocol(A, B) \& Internal(A) & Suspect(A) \\ 
 & 2 & Blocked(A, B) \& BlacklistService(A, B) \& Internal(A) & Suspect(A) \\ 
\hline
\multirow{2}{4em}{\centering{Spoofing}}
& 25 & Connect(A, B) \& PhysicalSpoofing(A, B) & Target(A), Target(B) \\
& 25 & Connect(A, B) \& LogicalSpoofing(A, B) & Target(A), Target(B) \\
\hline
\multirow{4}{4em}{\centering{Risky Incident}}
& 10 & Switch(A, S1) \& Switch(A, S2) \& (S1!=S2) & RiskyIncident(A) \\
& 100 & Switch(A, S1) \& Switch(A, S2) \& (S1!=S2) & HostMigration(A) \\
& 10 & SwitchFailure(S1) & RiskyIncident(S1) \\
& 10 & LinkFailure(S1, S2) & RiskyIncident(S1), RiskyIncident(S2) \\
\hline
\end{tabular}
\end{center}
\label{t:rca_rules}
\end{table*}

\textit{RCA} requires administrators to identify the initial attacker, detect infected hosts, and predict the attacker's next target by analyzing collected evidence. This involves extracting complex attack patterns from a provenance graph, a process that is often time-consuming and error-prone. To address this challenge, the \textit{root cause analyzer} leverages PSL~\cite{bach:jmlr17} to perform RCA by formalizing attack patterns as first-order logic. PSL relies on two key components: a \textit{dataset} representing known relationships and a \textit{ruleset} for inferring new relationships. In RCA, the dataset consists of previously generated SP features (Section~\ref{s:design-analyzer-feature}).

The \textit{root cause analyzer} requires \emph{RCA rules}, which are the first-order logic rules used in PSL. The RCA rules are defined as follows:  
\begin{equation} \label{e:rca_format} 
\textit{Weight} : \textit{SP}_1 \;\&\; \textit{SP}_2 \;\&\; \dots \;\&\; \textit{SP}_n \rightarrow \textit{Result},
\end{equation}  
where \textit{Weight} represents the significance of the rule and its impact on the resulting value, \textit{SP} denotes the SP features combined using logical conjunction (\&), and \textit{Result} indicates the inferred outcome, which is classified into three categories---\textit{Suspect}, \textit{Target}, and \textit{Risky Incident}---to represent the likelihood of a security event or condition occurring.

\begin{itemize}
    \item \textit{Suspect}: It represents a host that is either the initial attacker or has performed an attack. A higher suspect value indicates a greater likelihood that the host initiated or executed an attack on other hosts.
    \item \textit{Target}: It represents a host that has been infected or is at risk of being compromised. A high target value suggests a strong probability that the host is already infected or vulnerable to an attack.
    \item \textit{Risky Incident}: It refers to unexpected network changes, such as host migration or failures of links or switches, that pose potential security threats to the network.
\end{itemize}

The results derived through the RCA represent the probabilistic classification of each node, indicating the likelihood of its role within the attack sequence. In contrast to the \emph{state} assigned during the causality identification phase, which deterministically reflects the node’s status based on the observed causal relationships, the categories produced by RCA express the probabilistic characteristics of each node. Specifically, these values estimate the likelihood that a given node is the initial point of compromise or the target of an attack. Notably, during the RCA phase, a single host may simultaneously exhibit high probabilities of being both a \emph{suspect} and \emph{target}, a condition that typically arises when a compromised host is under the attacker’s control and is subsequently leveraged to launch further attacks against other hosts.

An example of the RCA rule is expressed as follows:
\begin{equation} \label{e:rca_example}
\begin{split}
5: \textit{Connect(A, B)} \;\&\; \textit{IDSTrojanActivity(A, B)} \\ \;\&\; \textit{Victim(B)} \rightarrow \textit{Suspect(A)}
\end{split}
\end{equation}  
This rule derives the result by combining multiple SP features, indicating that \texttt{host A} is classified as a \textit{suspect} when packet transmission occurs between \texttt{host A} and \texttt{host B}, which is identified as a Trojan activity, and \texttt{host B} is determined to be infected. If \texttt{host A} continues to compromise with additional hosts by exhibiting the same behavior, the likelihood of \texttt{host A} being a suspect increases. Table~\ref{t:rca_rules} summarizes the RCA rules currently supported by \ourtool{}. Administrators can also create and modify the RCA rules based on specific security policies and network environments.


\section{Evaluation}
\label{s:eval}

In this section, we will evaluate {\ourtool} for addressing the following key questions:  
\begin{itemize}  
    \item \textbf{Q1.} Can {\ourtool} effectively assist administrators in performing RCA in real-world intrusion scenarios? (Section~\ref{s:eval-attack})  
    \item \textbf{Q2.} What is the performance overhead introduced by {\ourtool}'s data collection and analysis? (Section~\ref{s:eval-per}) 
\end{itemize}

\subsection{Implementation}
\label{s:impl}

We implemented \ourtool{} using approximately 10,000 lines of Java code. The provenance collector (Section~\ref{s:design-collector}) was built using ONOS v1.8.4~\cite{berde2014onos}, which supports OpenFlow v1.3. Implemented as an ONOS application, it continuously monitors network changes, including flow rule creation/deletion, host migration, and topology modifications, to maintain updated network snapshots. For packet monitoring, dedicated monitoring flow rules are installed in a separate flow table to prevent interference with the existing flow rule tables used by other ONOS applications. This design allows network administrators to run any SDN application, including ONOS applications, without requiring modifications to integrate them with {\ourtool}.  

A provenance analyzer (Section~\ref{s:design-analyzer}) was implemented as a standalone application. The implementation leveraged several open-source software and libraries: OrientDB v2.2.23~\cite{orientdb} as the database, Blueprints v2.6~\cite{blueprints} for the database interface, Pcap4j v1.7~\cite{pcap4j} for packet capture, Kryonet v2.21~\cite{kryonet} as a NIO server/client network library, Snort v2.9.9~\cite{roesch1999snort} for intrusion detection and prevention, and Syslog4j v1.7~\cite{syslog4j} as a syslog server/client library. The root cause analyzer (Section~\ref{s:design-rca}) was implemented using PSL~\cite{psl}, a framework designed for developing probabilistic models. The root cause analyzer translates the RCA rules into PSL rules and integrates the collected provenance data to perform root cause analysis.



\subsection{Experimental Setup}
We performed our experiments on an Intel Hexa-core Xeon E5-1650 machine with 64GB of RAM using Mininet~\cite{mininet} to emulate the network topologies. For the performance evaluation, we utilized two Intel Quad-core i5-6600K machines with 16GB of RAM and four hardware SDN switches, consisting of two Pica8 P3290 switches and two Pica8 P3297 switches. Each machine ran on Ubuntu. The Xeon machine hosted LXD containers to execute {\ourtool} component instances, including OrientDB, Snort, and ONOS. We extended ONOS’s default forwarding application to support security forwarding, enabling the integration of security middleboxes along forwarding paths, as well as access control functionalities. Malicious payloads such as remote shell exploits and botnet messages were synthetically generated with signatures provided to the IDS (Snort) for detection and analysis.

\subsection{Attack Scenario Case Studies}
\label{s:eval-attack}  

Here, we will evaluate the effectiveness of \ourtool{} under two attack scenarios: (i) APT and (ii) botnet propagation.

\noindent\textbf{APT.} To evaluate the effectiveness of {\ourtool}, we demonstrate a scenario involving an APT, which is a prevalent attack in modern enterprise networks. APTs typically involve sophisticated techniques for infiltrating and compromising protected internal resources~\cite{symantec-apt}. For this demonstration, we present our motivating example---an information leakage attack described in Section~\ref{s:prob-motiv}---as the APT scenario to showcase {\ourtool}'s ability to construct a SP graph and perform RCA. We developed a script that generates the network topology shown in Figure~\ref{f:use_case_apt} using Mininet~\cite{mininet} and executed the exact attack sequences. Upon running the script, {\ourtool} collected and recorded the provenance data in its database and constructed a SP graph starting from the database server.

\begin{figure}
    \centerline{\includegraphics[width=1\linewidth]{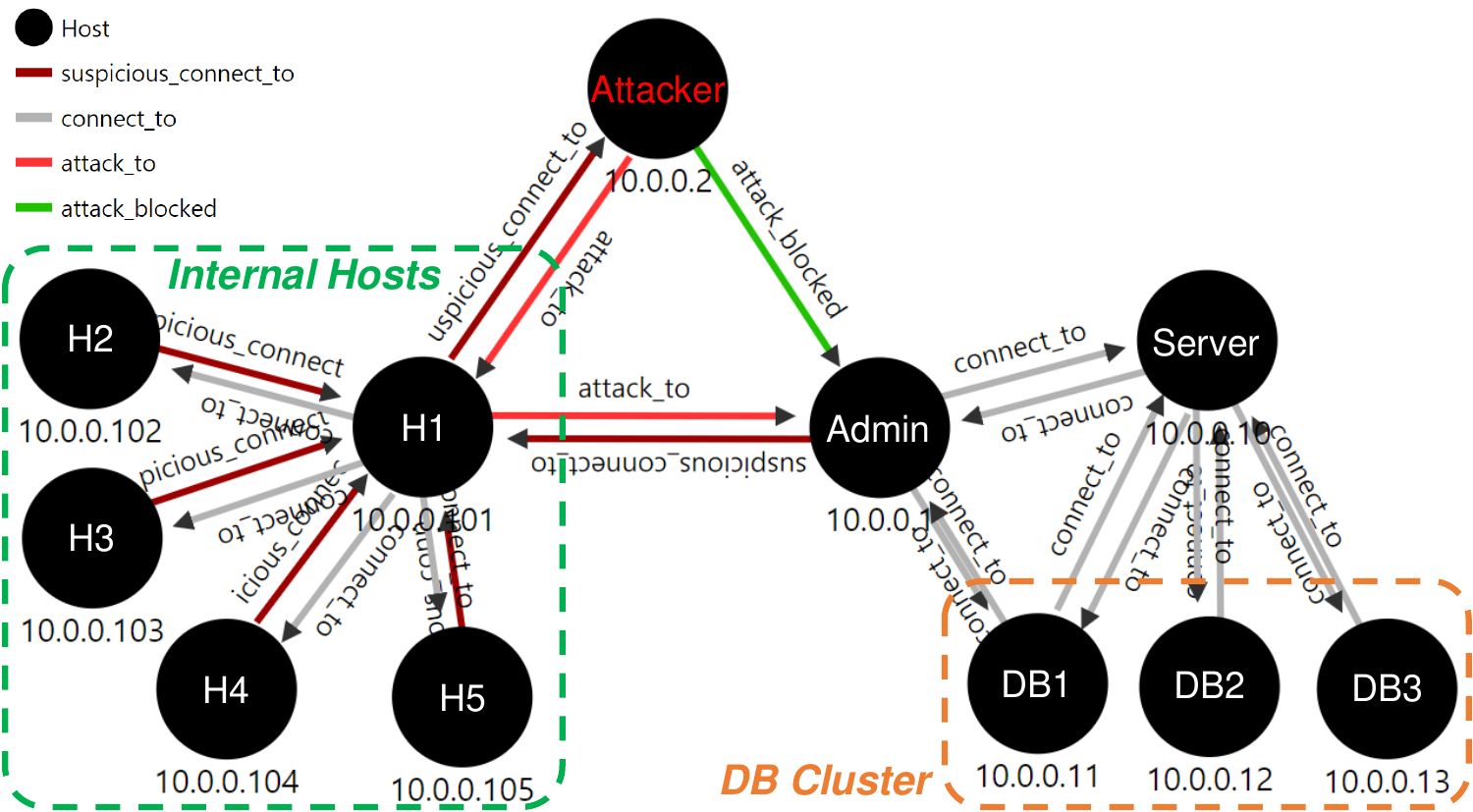}}
    \caption{Security provenance graph of the APT attack scenario in Figure~\ref{f:use_case_apt}. Some of the security causalities are omitted for visibility.}
    \label{f:eval_apt_graph}
\end{figure}

Figure~\ref{f:eval_apt_graph} presents the SP graph generated using {\ourtool}. The process begins by {\ourtool} retrieving evidence from one of the databases (DB1), identified by its IP address (10.0.0.11), to establish security causalities with other machines. DB1 has a \emph{connect\_to} relationship with the administrator machine, which in turn connects to the server for management. The server also maintains \emph{connect\_to} relationships with all the databases for operational purposes. The attacker host initially attempted to deliver remote exploitation to the administrator machine; however, the firewall successfully blocked this attempt. Consequently, {\ourtool} created an \emph{attack\_blocked} relationship between the attacker and the administrator machine, marking the attacker host as an \emph{attacker}. However, the internal host (H1) subsequently succeeded in exploiting the administrator machine. Consequently, {\ourtool} established an \emph{attack\_to} relationship between H1 and the administrator machine, classifying H1 as an \emph{attacker}.
\hmseo{
This analysis reveals that the attack propagated through four internal hosts, traversing from the attacker to H1, then to the administrator, the server, and finally to the database.}

Among the newly generated security causalities, \emph{attack\_to} has the highest priority, prompting \ourtool{} to select H1 as the next attack source (see Section~\ref{s:design-analyzer-graph}) and iteratively uncover additional security causalities. \ourtool{} further revealed that H1 conducted a port scan on other internal hosts, leading to the identification of eight security causalities, including \emph{connect\_to} and \emph{suspicious\_connect\_to}, linking H1 to internal hosts H2, H3, H4, and H5. In addition, two additional security causalities---\emph{attack\_to} and \emph{suspicious\_connect\_to}---were identified between the attacker and H1, confirming that H1 was initially compromised by the attacker. \\

\begin{figure}
\centering
   \includegraphics[width=.9\linewidth]{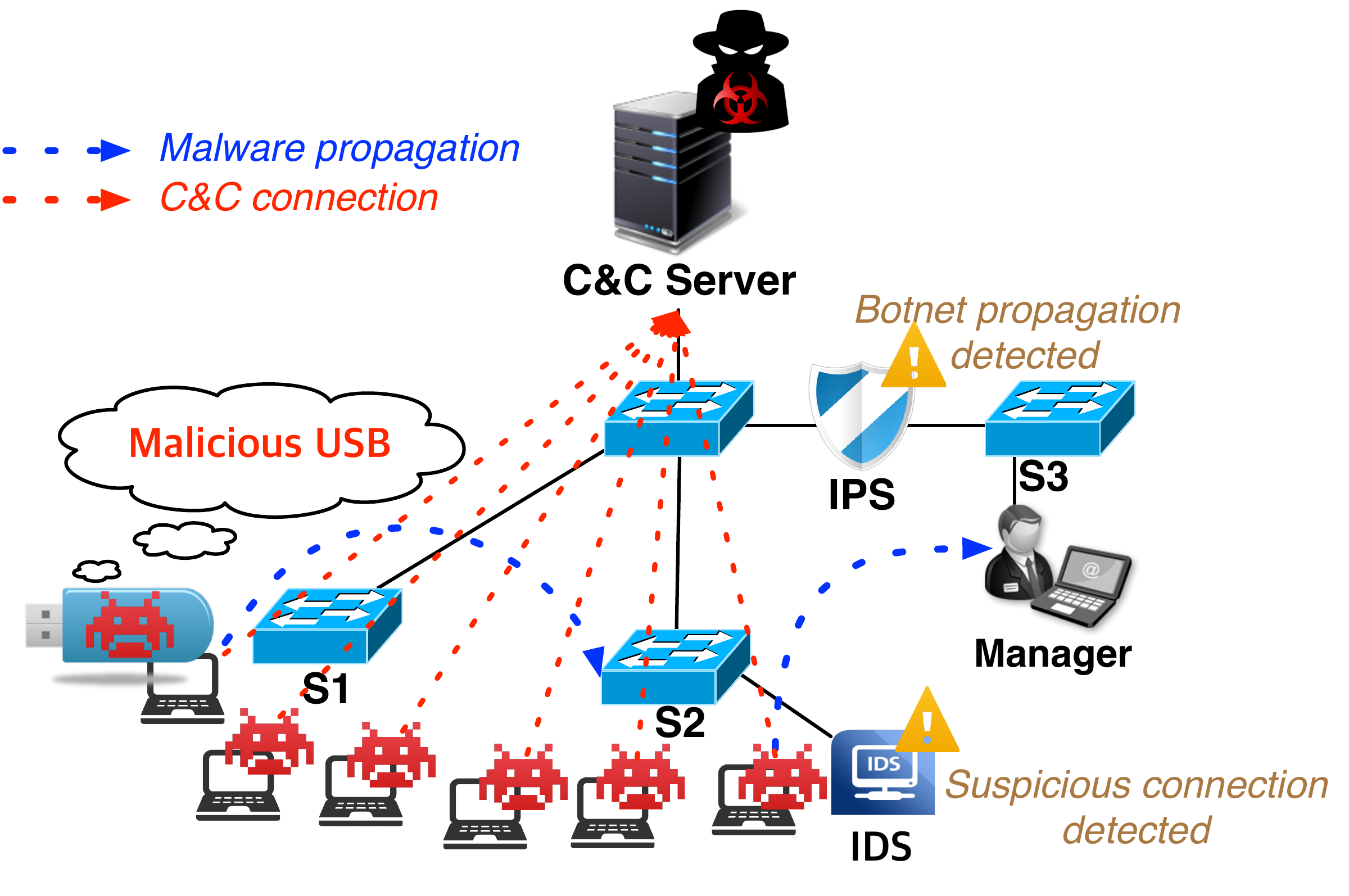}
    \caption{Scenario of enterprise botnet propagation by malware injection. One of the compromised hosts attempts to initiate access to the manager machine.}
    \label{f:use_case_botnet}
\end{figure}

\noindent\textbf{Botnet Propagation.}
Network attacks originating from internal hosts within an enterprise are challenging to detect and mitigate using conventional perimeter defense mechanisms. Although security middleboxes often generate frequent alerts, accurately assessing their relevance and severity remains a complex task. To evaluate the effectiveness of {\ourtool} in this context, we consider an enterprise attack scenario in which a botnet is propagated from an internal host that is inadvertently infected via a USB drive containing malware. Figure~\ref{f:use_case_botnet} illustrates a scenario in which the security manager receives alerts of blocked botnet propagation from an internal IPS and suspicious C\&C communication attempts detected by an internal IDS. Despite these alerts being generated by security systems, security administrators typically lack the contextual information required to accurately identify the attack origin or assess the security impact of such alerts.

\begin{figure}
\centering
   \includegraphics[width=.9\linewidth]{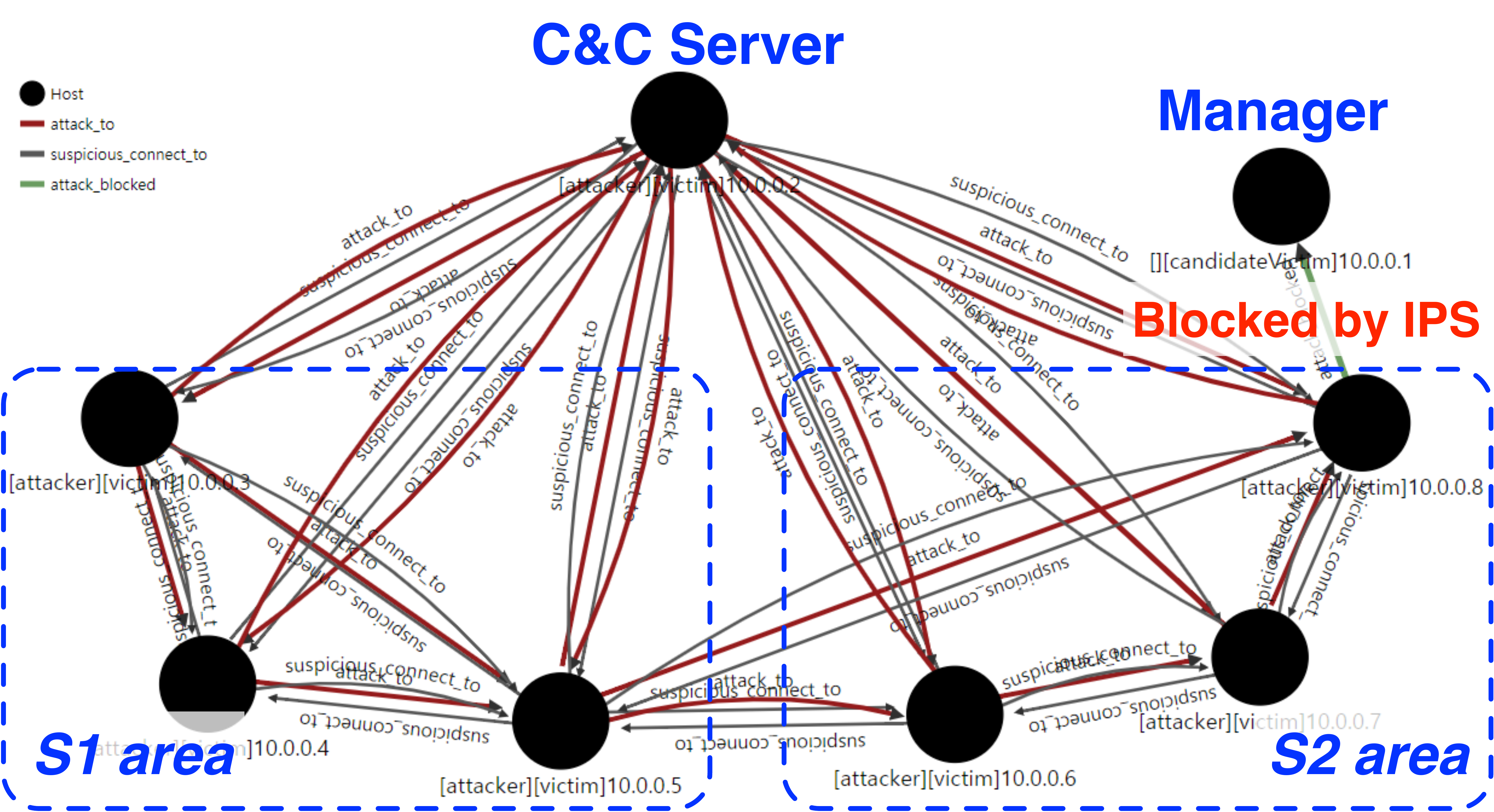}
    \caption{\ourtool{} reveals the trace of botnet pandemic from its threat propagation analysis.}
    \label{f:use_case_botnet_s}
\end{figure}

Figure~\ref{f:use_case_botnet_s} shows the attack-history graph generated using {\ourtool}. The botnet structure, consisting of a C\&C server and six infected hosts, emerges as a strongly connected component within the graph because of the proliferation of \textit{attack\_to} and \textit{suspicious\_connect\_to} edges derived from the collected attack evidence. Although the network traffic among hosts is complex and highly intertwined, the graph clearly shows that all communication paths converge toward the external C\&C server. Furthermore, it captures events in which a compromised host attempts to access the manager’s machine during an attack. This capability enables {\ourtool} to effectively visualize internal attacks that are otherwise difficult to detect, thereby providing a clear and comprehensive representation of attack propagation at the network level. In addition, the temporal information embedded in the edges can be leveraged to perform a RCA for deeper forensic investigations.

\begin{figure}
    \centering
    \subfloat[Probability of \textit{Suspect}.]{
        \includegraphics[width=0.5\linewidth]{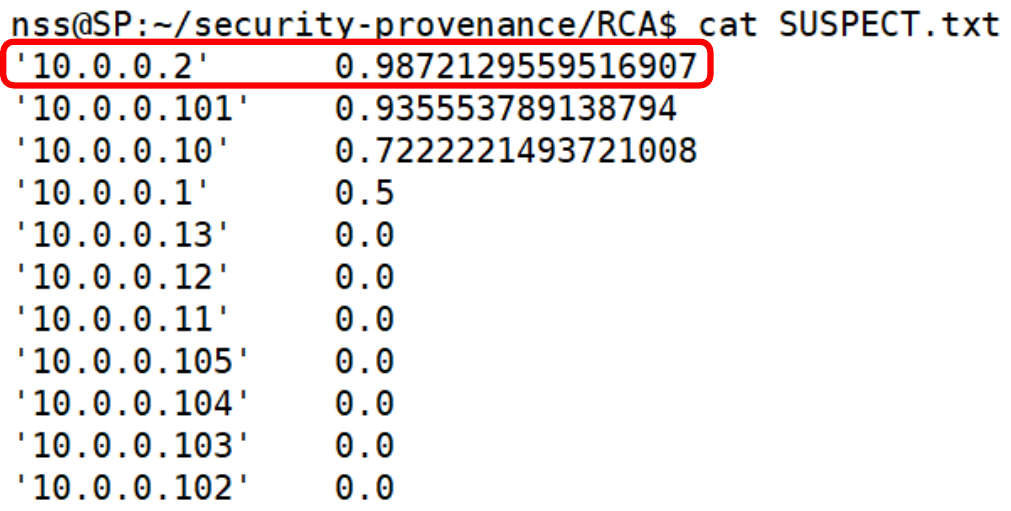}
        \label{fig:rca_eval1}
    }
    \subfloat[Probability of \textit{Target}.]{
        \includegraphics[width=0.5\linewidth]{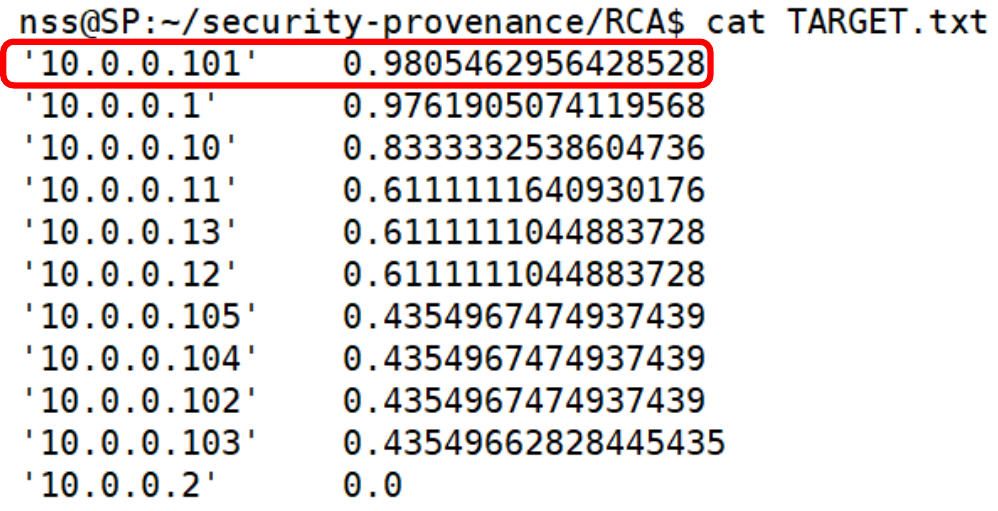}
        \label{fig:rca_eval2}
        
    }
    
    \subfloat[Probability of \textit{Risky Incident}.]{
        \includegraphics[width=0.5\linewidth]{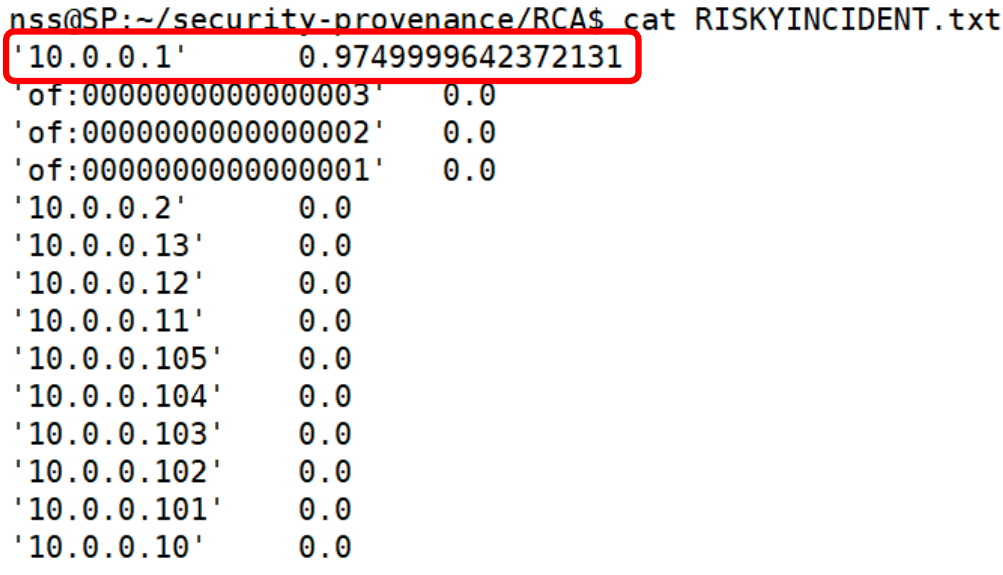}
        \label{fig:rca_eval3}
        
    }
    \caption{Results of applying RCA rules to the security provenance graph shown in Figure~\ref{f:eval_apt_graph}.}
    \label{fig:rca_eval}
\end{figure}

\subsection{Effectiveness of Root Cause Analysis}

We evaluated the RCA capability of \ourtool{} for the two attacks demonstrated.

\noindent
\textbf{APT.} After constructing the SP graph, {\ourtool} performed a RCA using PSL. The root cause analyzer (Section~\ref{s:design-rca}) extracted SP features from the constructed provenance graph shown in Figure~\ref{f:eval_apt_graph} and applied RCA rules to compute the probability of \emph{Suspect}, \emph{Target}, and \emph{Risky incident} for the attacker host (10.0.0.2), administrator host (10.0.0.1), and internal host H1 (10.0.0.101). In this experiment, all the database servers in the DB cluster (10.0.0.11, 10.0.0.12, and 10.0.0.13) were designated as \emph{victim} in the SP features (Table~\ref{t:sp_feature}).

\begin{figure}
    \centerline{\includegraphics[width=.9\linewidth]{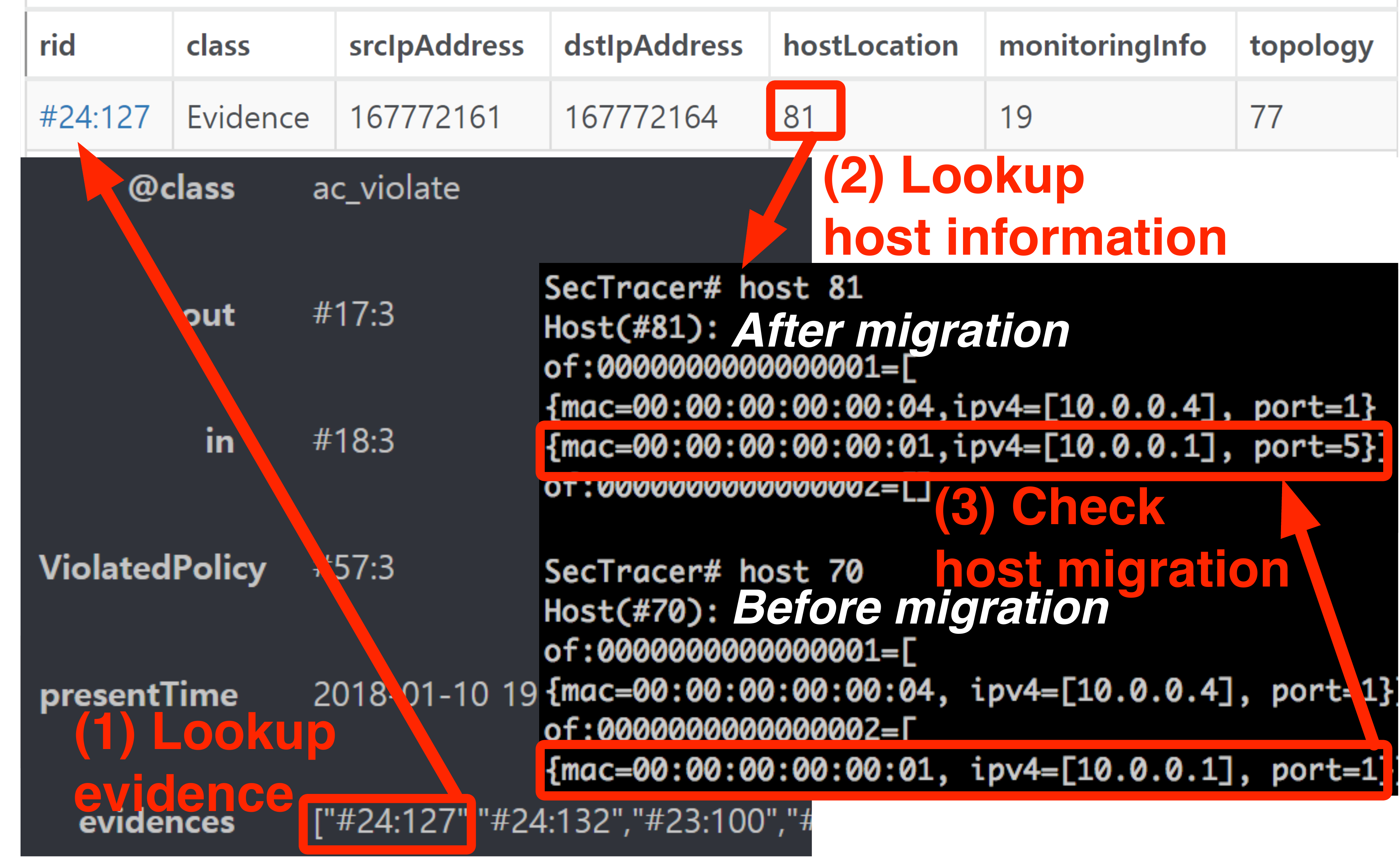}}
    \caption{Example of manual inspection with the \textit{ac\_violate} security causality evidences.}
    \label{f:use_case_acl_s}
\end{figure}

To infer the \emph{Suspect} probability, Figure~\ref{fig:rca_eval1} shows that the attacker host (10.0.0.2) and the internal host H1 (10.0.0.101) have the highest suspicion levels, with probabilities of 98.7\% and 93.6\%, respectively. Additionally, the server (10.0.0.10) and administrator machine (10.0.0.1) are identified as suspicious, with probabilities of 72\% and 50\%, respectively. This is because both machines have evidence marked as \emph{reached} on their database servers, which are designated as \emph{ victim}. Notably, server (10.0.0.10) has a higher suspect value than the administrator machine (10.0.0.1) because of its access to a greater number of database servers, which resulted in more causalities.  

To infer the \emph{Target} probability, Figure~\ref{fig:rca_eval2} shows that the internal host H1 (10.0.0.101) and administrator machine (10.0.0.1) have the highest probabilities of being targeted at 98\% and 97.6\%, respectively. In addition, server (10.0.0.10) has a high probability (83.3\%) because it is accessible to all database servers, making it a likely target for attackers. All database servers (10.0.0.11, 10.0.0.12, and 10.0.0.13) have a 61.1\% probability of being targeted because they are designated as \emph{ victim}. Furthermore, all other internal hosts (10.0.0.102, 10.0.0.103, 10.0.0.104, and 10.0.0.105) have a 43.5\% probability because they were scanned by H1 but did not experience further malicious activity.  

Finally, to infer the \emph{Risky Incident} probability, Figure~\ref{fig:rca_eval3} shows that the administrator machine (10.0.0.1) has the highest probability of 97.5\%. An in-depth analysis was conducted to determine this cause. To support administrator inspection, \ourtool{} provides both a GUI dashboard and a CLI tool. Figure~\ref{f:use_case_acl_s} illustrates this process, which enables administrators to perform RCA effectively. During the investigation, an administrator detected an \texttt{ac\_violation} causality originating from the SDN switch with DPID \emph{of:0000000000000001}. However, further analysis revealed that the machine’s current location was at a different SDN switch from the DPID \emph{of:0000000000000002}, confirming that a host has migrated. This migration event could lead to access-control policy violations, thereby exposing potential security risks. \\

\begin{figure}
\centering
   \includegraphics[width=.9\linewidth]{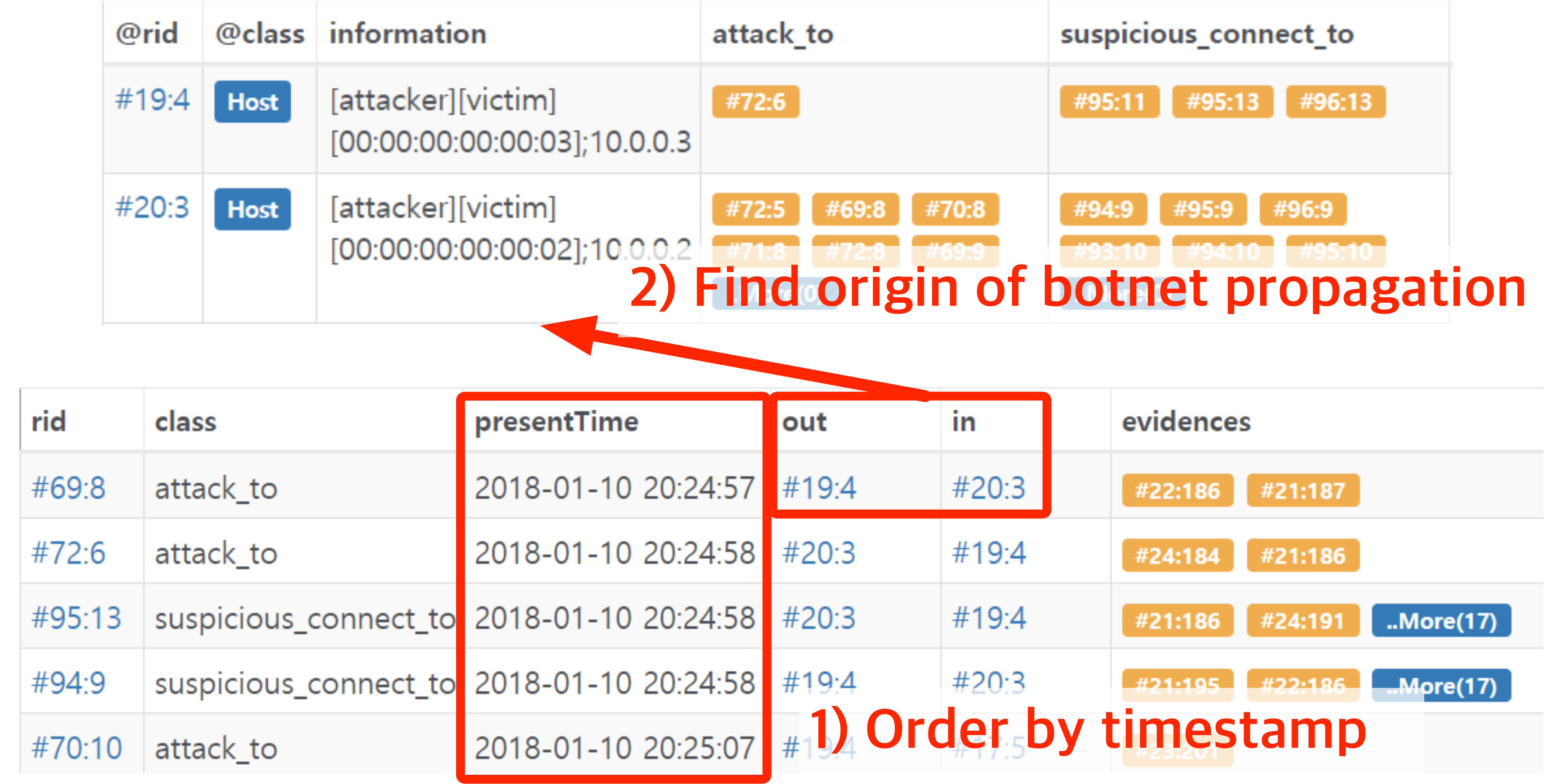}
    \caption{Chronological order of generated security relationships in {\ourtool} dashboard GUI. The head of the list corresponds to the attack origin.}
    \label{f:use_case_botnet_sql}
\end{figure}

\noindent
\textbf{Botnet Propagation.}
In the botnet propagation scenario, identifying the root cause is comparatively more straightforward than in the APT scenario. As all connections in a botnet communicate directly with the C\&C server, the root cause can often be determined without performing a full RCA, which relies solely on attack forensics to pinpoint the origin. The origin of the attack can be inferred from the chronological ordering of the identified security relationships. Figure~\ref{f:use_case_botnet_sql} shows the {\ourtool} dashboard GUI, which lists security relationships in timestamp order. Tracing the initial entry in this sequence reveals the primary point of compromise, namely, the host infected via the malicious USB drive.

\subsection{Performance Benchmarks}
\label{s:eval-per}

We evaluated {\ourtool} in terms of system overhead and performance. To assess the overhead of {\ourtool}'s provenance collection, we used \texttt{iperf3}~\cite{iperf} and \texttt{ping}~\cite{ping} to measure its impact on network performance. Additionally, we assessed the processing time of the submodules in evidence generation and graph construction to comprehend their performance bottlenecks. \\

\begin{figure}
    \centerline{\includegraphics[width=.95\linewidth]{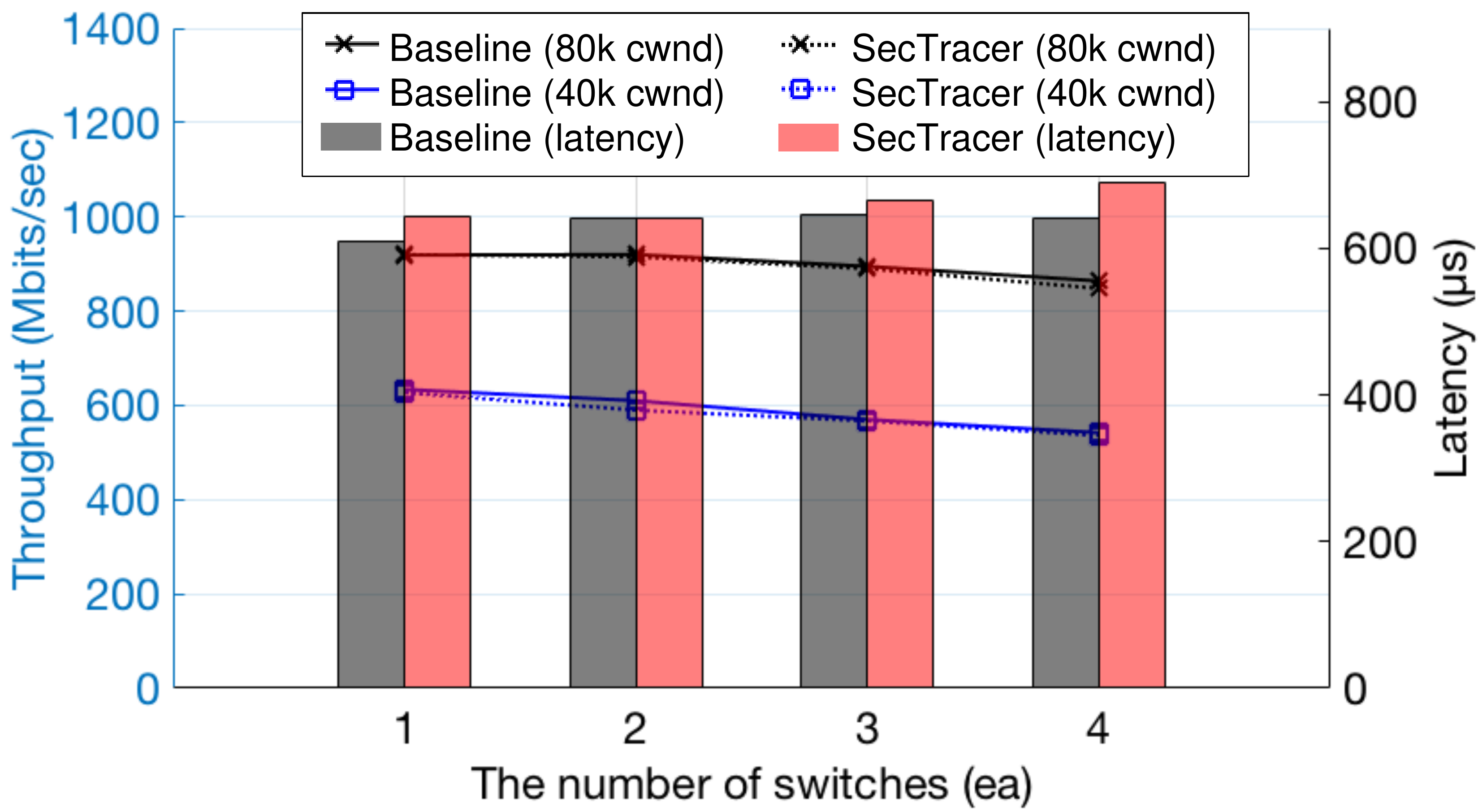}}
    \caption{End-to-end latency and throughput with/without {\ourtool}.}
    \label{f:overhead_latency_throughput}
    \vspace{-0.1in}
\end{figure}

\noindent\textbf{Network Overhead of \ourtool{}.} We evaluated the end-to-end latency and throughput with and without provenance collection of {\ourtool} to assess its overhead. Our experimental environment consisted of two Intel i5 machines and four Pica8 SDN switches, and we configured a linear topology by incrementally adding the switches. Forwarding rules were installed to enable packet delivery between the two hosts, whereas monitoring rules were issued by {\ourtool} to the edge switches (or a single switch). We used iPerf3 to measure TCP throughput with varying window sizes and ping to evaluate the latency. Figure~\ref{f:overhead_latency_throughput} shows the evaluation results. As the number of switches increases, the throughput decreases slightly in both the baseline and {\ourtool} cases. However, {\ourtool} introduces less than 1\% overhead when using TCP window sizes (cwnd) of 40k and 80k. In terms of latency, {\ourtool} exhibits negligible overhead compared with the baseline, demonstrating its efficiency in provenance collection without significant performance degradation. \\

\begin{table}
\scriptsize
\caption{Elapsed time ($\mu$s) of the evidence generation. The star-notation (\textsuperscript{*}) indicates cached results. ({\em{I}}:~Integrity, {\em{R}}:~Reachability, {\em{PV}}:~Policy Violation, {\em{SP}}:~Security Policy, {\em{SD}}:~Security Device)}
\begin{center}
\begin{tabular}{l|c|c|c|c|c|c}
 & \bf{I} & \bf{R} & \bf{PV I} & \bf{PV II} & \bf{SD} & \bf{Total}\\
\hline 
\hline 
\bf{\textit{Benign}} & 2.50 & 4.45 & 134.31  & 127.49  & - & 268.75 \\
 &  &  & (0.03)\textsuperscript{*} & (0.14)\textsuperscript{*} &  & \\
\hline 
\bf{\textit{Malicious}} & 2.50 & 4.45 & 134.31 & 127.49 & 811.58 & 1,080.33 \\
 &  &  &  (0.03)\textsuperscript{*} &  (0.14)\textsuperscript{*} &  & \\
\hline 
\end{tabular}
\end{center}
\label{t:evidence_generation}
\vspace{-0.15in}
\end{table}


\noindent\textbf{Microbenchmark of Evidence Generation.}
To evaluate the performance overhead of provenance collection, we measured the processing time of each submodule in the evidence generation process of \ ourtool{}, including integrity, reachability, and policy violation checks (Section~\ref{s:design-collector-evidence}). Table~\ref{t:evidence_generation} presents the microbenchmark results for both benign and malicious cases. In both scenarios, integrity ({\em{I}}) and reachability ({\em{R}}) were completed within 5~$\mu s$. We assessed two cases of policy violation detection: (i) deploying 10 security routing rules (\emph{PV I}) and (ii) deploying 10 access control rules (\emph{PV II}). Checking for \emph{PV I} and \emph{PV II} violations requires 134.31~$\mu s$ and 127.49~$\mu s$, respectively, as the processing time increases linearly with the number of rules. The primary bottleneck stems from \ourtool{} retrieving provenance data from a database through rule matching, which can be optimized by caching frequently accessed rules. With caching, the processing times for \emph{PV I} and \emph{PV II} are reduced to 0.03~$\mu s$ and 0.14~$\mu s$, respectively, achieving a 99.9\% reduction. In addition, we evaluated the processing time when \ourtool{} relies on existing security devices (\emph{SD}), which requires parsing IDS logs to obtain evidence. Because security devices generate logs only upon detecting malicious activity, we measured this case separately. The results indicate that leveraging security devices accounts for over 75\% of the total processing time because of the string parsing overhead required to interpret IDS logs. \\

\noindent\textbf{Microbenchmark of Graph Construction.} We evaluated the graph construction performance by measuring the processing time of each submodule. Using a similar experimental setup (i.e., an APT attack) on a Xeon machine running OrientDB and Snort, we obtained the results listed in Table~\ref{t:overall_time_apt}. Overall, \ourtool{} completed the graph construction in approximately 31 s. It generated 32,155 pieces of evidence (\emph{EG}) in approximately 1.57 s, accounting for 4.9\% of the total processing time. In addition, \ourtool{} produced 32,155 security causalities (\emph{SR}) in only 0.073 s and performed 32,114 state-transition operations (\emph{ST}) in 0.006 s. The most time-consuming step was the database operation (\emph{DB Ops}), which involved 39 read/write executions to retrieve packets and store vertices/edges for the SP graph. These operations accounted for approximately 94.75\% of the total processing time. To enhance the performance of \ourtool{}, we plan to replace OrientDB with a high-performance database and leverage distributed computing environments~\cite{zaharia2010spark} to efficiently distribute workloads.

\begin{table}
\scriptsize
\caption{Processing time for the attack graph of the APT attack scenario.  (\emph{EG}:~Evidence Generation, {\em{SC}}:~Security Causality, {\em{ST}}:~State Transition, \emph{DB Ops}:~DB Operations)}
\begin{center}
\begin{tabular}{l|c|c|c|c}
 & \bf{EG} & \bf{SC} & \bf{ST} & \bf{DB Ops}\\
\hline \hline
\bf{\textit{\# of Operations}} & 32,155 & 32,155 & 32,114 & 39\\
\hline 
\bf{\textit{Per-Operation Time ($\mu$s)}} & 48.75 & 2.27 & 0.19 & 763,272.21\\
\hline 
\bf{\textit{Total Time (s)}} & 1.569 & 0.073 & 0.006 & 29.767\\
\hline 
\bf{\textit{Ratio (\%)}} & 4.994 & 0.232 & 0.019 & 94.754\\
\end{tabular}
\end{center}
\label{t:overall_time_apt}
\vspace{-0.1in}
\end{table}


\section{Discussion and Limitations}
\label{s:discussion}

In this section, we discuss the additional design considerations and limitations of \ourtool{}. \\

\noindent
\textbf{Real-time Attack Detection and Response.}
Because \ourtool{} is primarily designed for forensic analysis to uncover the root causes of APTs in enterprise networks, its real-time detection and response capabilities are limited. However, this limitation can be mitigated by integrating traditional firewalls or NIPS to deploy filtering rules for suspicious hosts identified through RCA. Furthermore, in SDN-enabled networks---as assumed in Section~\ref{s:prob-threat}---such responses can be efficiently implemented by installing blocking rules directly on SDN switches. \\

\noindent
\textbf{Storage Optimization for Security Logging.}
A security snapshot is generated only when changes occur in the corresponding policies or network configurations. As these elements do not change frequently, the overall size of the logging data remains relatively small. Furthermore, evidence records reference only the ID of the most recent snapshot, rather than the full snapshot, ensuring that data size does not become a concern. In addition, employing a delta-based storage approach, where only the differences between snapshots are stored, can further minimize the volume of logging data. Similarly, network packets can be aggregated using an optimized process, where redundant packets are deduplicated and only changes or differences are recorded before storage. Beyond these methods, various network log retention strategies can be applied to optimize storage efficiency and further reduce the size of the logging dataset. \\

\noindent
\textbf{Temporal-aware Causality in the Graph.}  
Although \ourtool{} embeds timestamps in security causality edges, as shown in Figure~\ref{f:provenance_graph_overview}, it does not enforce the temporal order between vertices. This implies that an edge is created between entities when an incident occurs, regardless of their chronological order. To reconstruct an attack timeline, administrators must refer to the timestamps embedded in each causality edge. Although this design requires additional effort from administrators, it is motivated by the need to avoid the \emph{dependency explosion} problem, where the provenance graph grows significantly in size if full temporal ordering among vertices is enforced. Thus, our approach prioritizes graph efficiency while preserving temporal information within the edges. \\

\noindent
\textbf{False Positives in RCA.}  
Despite our tamper-resistant provenance collection (Section~\ref{s:design-collector-evidence}), \ourtool{} may incorrectly identify the root causes owing to inaccurate provenance data or labeling errors. Such inaccuracies stem from the use of the indirect causality mechanisms described in Section~\ref{s:design-graph-causality}, including \texttt{risky\_connect\_to}, \texttt{implicit\_violate\_ac}, and \texttt{implicit\_violate\_sp}, which infer potential causality rather than rely on concrete evidence. However, the goal of \ourtool{} is not to guarantee precise causal inference but rather to highlight potential threats based on predefined rules, enabling administrators to initiate manual investigations more efficiently. This approach helps reduce the burden of analyzing complex provenance graphs. \\

\noindent
\textbf{Comparison with Host-level Provenance Graphs.}  
\ourtool{} focuses on network-level SP to identify the root causes of APTs that span multiple hosts. Consequently, it does not provide the process-, user-, or file-level granularity offered by traditional host-level provenance graphs. We argue that RCA at the network level is critical in enterprise environments. Nevertheless, to track intrahost activities, \ourtool{} can be integrated with host-level provenance systems to provide more fine-grained visibility. In this sense, \ourtool{} is complementary to, rather than a replacement for, host-level provenance graphs.


\section{Conclusion}
Network intrusions in enterprise networks are becoming increasingly sophisticated, making their detection and analysis challenging. To address this problem, we introduce \ourtool{}, a novel network-level provenance-based intrusion detection system that enables administrators to gain a clear understanding of operational scenarios and attack origins. \ourtool{} systematically collects provenance data, constructs provenance graphs, and performs RCA using probabilistic models. Our evaluation demonstrates that \ourtool{} effectively tracks the detailed operations of APT attacks and accurately identifies their root causes with minimal system overhead. These findings suggest that our approach is well-suited to securing modern network infrastructures, even in complex environments.

\printcredits

\bibliographystyle{unsrt}

\bibliography{references}


\newpage

\bio{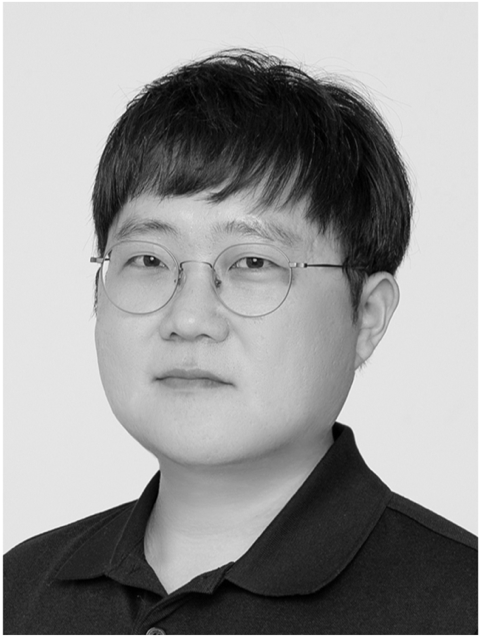}
\textbf{Seunghyeon Lee} is the Product Director of the AI Product Department and Co-Founder at S2W. He received his Ph.D. degree in Information Security from KAIST. His research interests include ontology modeling, knowledge graph construction, and intelligent information systems. His current work focuses on applying these research areas to develop AI products that address complex real-world problems.

\endbio

\vspace{1in}

\bio{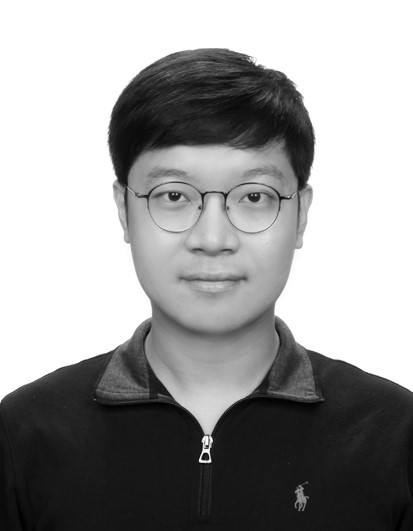}
\textbf{Hyunmin Seo} is a Ph.D. student in the School of Electrical Engineering at KAIST. 
He received a B.S. degree in Electrical Engineering from KAIST. 
He received his M.S. degree in Electrical Engineering from KAIST. 
His research interests include programmable network data planes, cyberattacks, and cloud security.
\endbio

\vspace{1in}

\bio{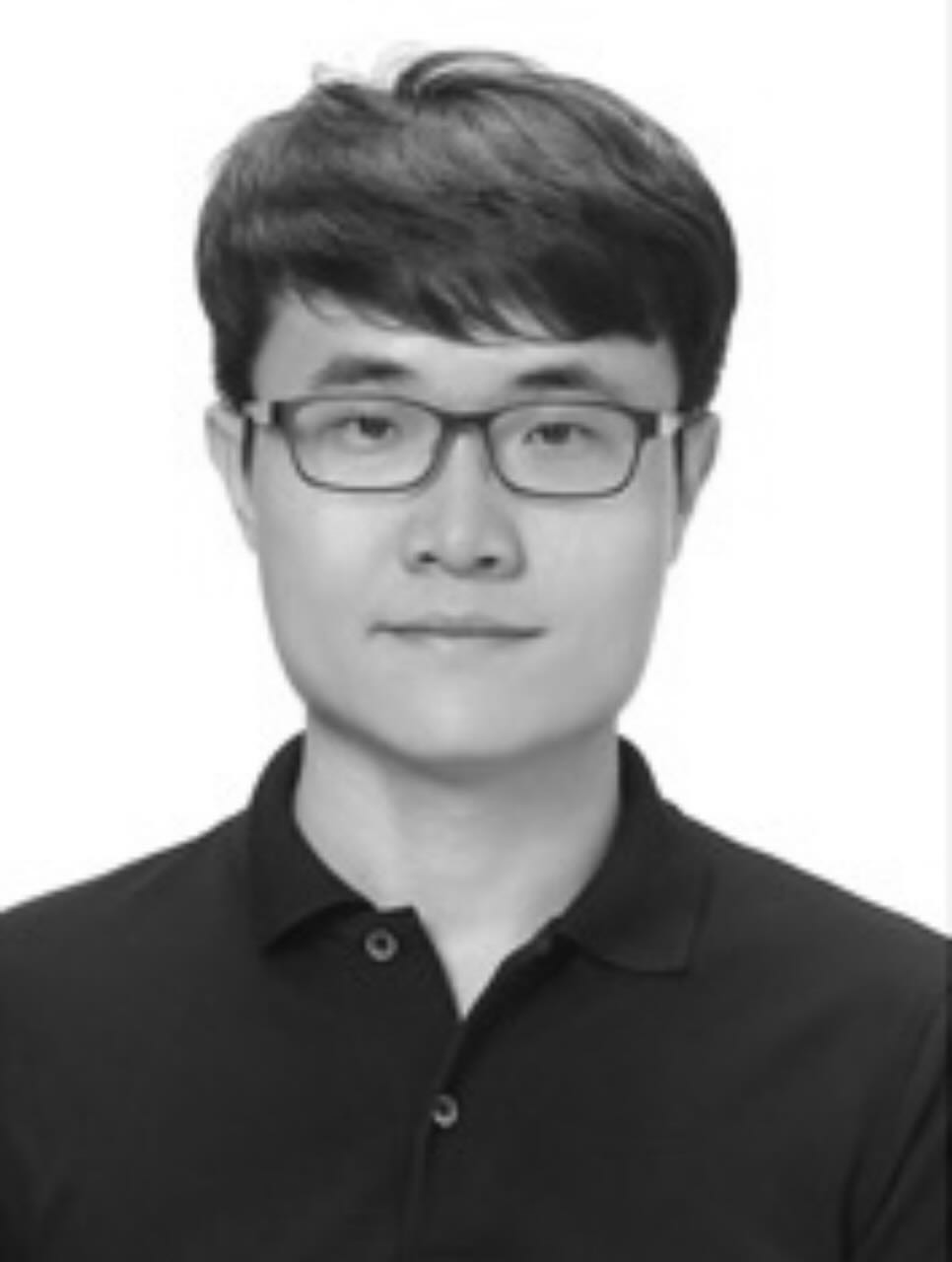}
\textbf{Hwangjo Heo} is a Principal Researcher at the Electronics and Telecommunications Research Institute (ETRI). He earned his Ph.D from KAIST and his Master’s degree from Purdue University. His research interests include the security of computer systems, encompassing blockchain, computer networks, and artificial intelligence systems. 
\endbio

\vspace{1in}

\bio{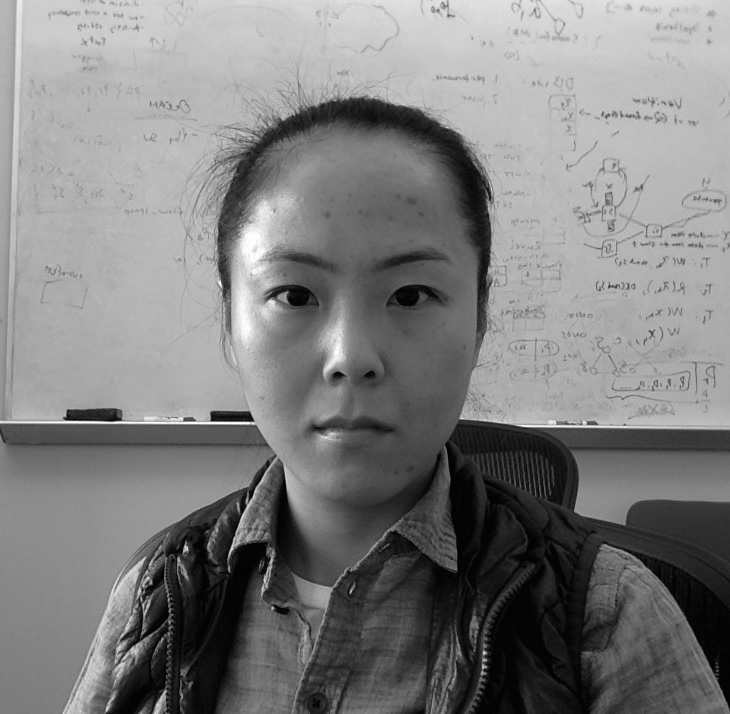}
\textbf{Anduo Wang} is an Associate Professor at Temple University, where she specializes in improving network state management using formal methods, databases, knowledge representation and reasoning, and logic programming. She received her Ph.D. from the University of Pennsylvania in 2013.
\endbio

\newpage

\vspace{1in}

\bio{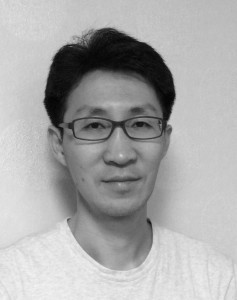}
\textbf{Seungwon Shin} is an Associate Professor in the School of Electrical Engineering at KAIST and an Executive Vice President at Samsung Electronics. He received his Ph.D. degree in Computer Engineering from the Electrical and Computer Engineering Department, Texas A\&M University, and his M.S. and B. S. degrees from KAIST, both in Electrical and Computer Engineering. His research interests include software-defined networking security, dark web analysis, and cyber threat intelligence.
\endbio

\vspace{1in}

\bio{author_photo/bio_kim}
\textbf{Jinwoo Kim} is an Assistant Professor in the School of Software at Kwangwoon University, Seoul, Republic of Korea. He received his Ph.D. degree from the School of Electrical Engineering at KAIST, his M.S. degree from the Graduate School of Information Security at KAIST, and his B.S. degree from Chungnam National University in Computer Science and Engineering. His research focuses on security and privacy issues with software-defined networks, cloud systems, and VR.
\endbio

\end{document}